\documentclass[superscriptaddress,dvipsnames,nofootinbib,amsmath,amssymb,prd,notitlepage]{revtex4-1}

\usepackage{subfig}
\usepackage[a4paper, left=20mm,right=20mm,top=20mm,bottom=20mm]{geometry}
\usepackage[titletoc,toc,title]{appendix}

\usepackage{mathtools}
\usepackage{tensor}
\usepackage{amsmath, amsfonts,bm}
\usepackage{derivative}

\usepackage{bm}
\usepackage[utf8]{inputenc}
\usepackage[normalem]{ulem}
\usepackage{nicefrac}

\usepackage{suffix}

\usepackage{amsmath, amsfonts,bm}
\usepackage{xcolor}
\usepackage{graphicx}
\definecolor{darkerred}{rgb}{0.9, 0.17, 0.31}
\usepackage{subfig}

\usepackage[colorlinks=true,hyperfootnotes=true, linkcolor=darkerred, citecolor=cyan, urlcolor=magenta]{hyperref}

\DeclareMathAlphabet{\mathup}{OT1}{\familydefault}{m}{n}

\def\dd{\mathrm{d}}

\newcommand{\be}{\begin{equation}} 
\newcommand{\ee}{\end{equation}}

\usepackage[capitalise]{cleveref}

\usepackage{array}
\newcommand{\PreserveBackslash}[1]{\let\temp=\\#1\let\\=\temp}
\newcolumntype{C}[1]{>{\PreserveBackslash\centering}p{#1}}
\newcolumntype{R}[1]{>{\PreserveBackslash\raggedleft}p{#1}}
\newcolumntype{L}[1]{>{\PreserveBackslash\raggedright}p{#1}}

\begin{document}

\title{Interacting dark sector from intrinsic entropy couplings}

\author{Erik Jensko}
\email{erik.jensko@ucl.ac.uk}
\affiliation{Department of Mathematics, University College London,
Gower Street, London WC1E 6BT, UK}

\author{Elsa M. Teixeira}
\email{elsa.teixeira@umontpellier.fr}
\affiliation{Laboratoire Univers \& Particules de Montpellier,
CNRS \& Université de Montpellier (UMR-5299), 34095 Montpellier, France}

\author{Vivian Poulin}
\email{vivian.poulin@umontpellier.fr}
\affiliation{Laboratoire Univers \& Particules de Montpellier,
CNRS \& Université de Montpellier (UMR-5299), 34095 Montpellier, France}

\begin{abstract}
We introduce a new class of interacting dark sector models that couple the intrinsic entropy of dark matter to scalar field dark energy. Using the Lagrangian formulation for relativistic perfect fluids, we construct consistent covariant actions that incorporate algebraic and derivative entropy couplings. These interactions leave the expansion history unchanged, rendering the background cosmology indistinguishable from $\Lambda$CDM or uncoupled quintessence.
At the level of cosmological perturbations, the entropy couplings generate scale-dependent modifications to the dark matter Euler equation, while the continuity equation remains unaltered at linear order. 
The resulting interactions correspond to a pure-momentum exchange within the dark sector. We show that intrinsic entropy perturbations can carry primordial scale dependence, and non-minimal couplings can lead to a scale-dependent  suppression or enhancement of structure growth.
Finally, we demonstrate that these models are generically compatible with current Cosmic Microwave Background observations, while inducing distinctive signatures in large-scale structure. 
The framework provides a theoretically well-motivated and observationally viable extension to the standard cosmological model, opening new directions to explore novel interactions in the dark sector.

\end{abstract}

\maketitle

{\hypersetup{linkcolor=black}
\makeatletter
\def\l@subsection#1#2{}
\def\l@subsubsection#1#2{}
\makeatother
\tableofcontents
}

\section{Introduction}

The standard model of cosmology, known as the $\Lambda$CDM model, successfully describes a wide range of cosmological observations, from measurements of the anisotropies in the cosmic microwave background (CMB)  to the formation and evolution of large-scale structure in the Universe. Within this model, dark energy, the driving component of the accelerated expansion of the Universe \cite{SupernovaCosmologyProject:1998vns,SupernovaSearchTeam:1998fmf}, is identified with the cosmological constant $\Lambda$, a non-diluting energy density contribution. Additionally, dark matter provides the gravitational scaffolding necessary for structure formation and is modelled as cold dark matter (CDM) in the standard framework, a weakly interacting, unknown form of non-baryonic matter. 
Together, these two components form the dark sector.
Despite its remarkable successes, the $\Lambda$CDM model has faced growing challenges over the past couple of decades. 
On the theoretical side, the precise nature of the dark sector is still unknown, and the value of the cosmological constant is hard to reconcile with its interpretation as the vacuum energy in quantum field theory \cite{Weinberg:1988cp,Martin:2012bt}.
On the observational side, discrepancies in independent measurements of the present-day expansion rate ($H_0$) \cite{Riess:2021jrx,H0DN:2025lyy} and the amplitude of matter clustering ($\sigma_8$) \cite{Heymans:2020gsg,Wright:2025xka,DES:2025tna} may point to missing ingredients or unaccounted-for physical processes in the Standard Model. These unresolved cosmic tensions
have become a central focus of modern cosmology and have attracted significant attention in recent years \cite{Verde:2019ivm,Abdalla:2022yfr,CosmoVerseNetwork:2025alb}.

In particular, models with interactions between dark matter and dark energy have been widely explored as extensions
of $\Lambda$CDM (see Ref.~\cite{Wang:2024vmw} for a review). Indeed, such interactions are natural and would be expected to occur in nature unless prevented by additional symmetries \cite{Carroll:1998zi,Amendola:1999er,Wang:2016lxa}. A wide variety of couplings, typically introduced in the context of scalar fields or within effective fluid descriptions, have been proposed to modify the cosmic expansion history and to enhance or suppress the growth of structure \cite{Aoki:2025bmj,Poulot:2024sex,vandeBruck:2022xbk,Bettoni:2013diz,Zumalacarregui:2013pma,Gubitosi:2012hu,Frusciante:2019xia,Smith:2024ibv,vandeBruck:2015ida,Linton:2021cgd,Amendola:2020ldb,Barros:2019rdv,VanDeBruck:2017mua,Bansal:2024bbb}.
Consequently, interacting dark sector models have been extensively studied as possible explanations of the observed cosmic tensions and as potential resolutions to small-scale structure issues \cite{DiValentino:2017iww,Poulin:2018zxs,DiValentino:2019jae,Teixeira:2024qmw,DiValentino:2019ffd,Gomez-Valent:2020mqn,Pourtsidou:2016ico,Lucca:2020zjb,Becker:2020hzj,Castello:2023zjr,Tsedrik:2025jdv,Goh:2024exx,Teixeira:2022sjr,Bansal:2025usn}. Of particular recent interest are the pure-momentum transfer models (e.g., modelling dark matter elastic scattering), which evade tight background cosmological constraints while leading to measurable effects in observables sensitive to matter clustering \cite{Pourtsidou:2013nha,Linton:2021cgd,BeltranJimenez:2022irm,Poulin:2022sgp,Kumar:2017bpv,Baldi:2014ica,Simpson:2010vh,Asghari:2019qld,BeltranJimenez:2020qdu,BeltranJimenez:2021wbq}. Aside from the well-studied Type-3 models of Ref.~\cite{Pourtsidou:2013nha} and velocity-entrainment models \cite{BeltranJimenez:2020qdu,BeltranJimenez:2021wbq}, most other pure-momentum exchange models are phenomenological, highlighting the lack of compelling fundamental frameworks capable of producing momentum exchange derived from Lagrangian principles.

In this paper, we point out that couplings between the \textit{intrinsic entropy} of dark matter fluids and scalar field dark energy have been overlooked in previous work. 
Intrinsic entropy perturbations arise in the effective fluid description of many dark energy or inflationary models \cite{Bean:2003fb,Hu:2004kh,Cicoli:2021itv,Arjona:2020yum}, and are closely related to relative entropy (isocurvature) perturbations
\cite{Wands:2000dp,Malik:2002jb,Malik:2004tf,Bartolo:2003ad,Arjona:2020yum,Bemfica:2020zjp,Kopp:2016mhm,Valiviita:2008iv,Hu:1998kj,Thomas:2016iav}. 
In this context, we present a new formalism to describe \textit{entropic interactions} in the dark sector at the effective fluid level by introducing intrinsic entropy as a new coupled degree of freedom. Importantly, we show that entropy couplings directly lead to pure-momentum transfer in the dark sector.
This is achieved by utilising the Lagrangian formulation pioneered by Schutz and Sorkin Ref.~\cite{Schutz:1977df} and Brown \cite{Brown:1992kc} to describe perfect cosmological fluids, later applied to non-minimally coupled models such as in Refs.~\cite{Pourtsidou:2013nha,Skordis:2015yra,Koivisto:2015qua,Boehmer:2015kta,Boehmer:2015sha,Dutta:2017kch,Boehmer:2024rqk,Ashi:2025dba,Boehmer:2026syd}.  
Given that many phenomenological interacting models suffer instabilities when couplings are added by hand \cite{Koivisto:2005nr,Valiviita:2008iv,Bean:2008ac,Gavela:2009cy,Martinelli:2019dau}, our Lagrangian treatment provides a consistent way to avoid these issues, with perturbations unambiguously derived from variational principles. Moreover, an additional advantage of this formulation is its compatibility with standard stability techniques that require a Lagrangian and the close connection with effective field theory (EFT) approaches \cite{Ballesteros:2012kv,Aoki:2025bmj}.

For cosmological applications, we specialise to an \textit{entropic-CDM} fluid, which is barotropic and adiabatic with vanishing pressure and sound speed $w=c_s^2=0$. This simplifies the analysis since the non-adiabatic pressure perturbations vanish, in contrast to past studies of entropy-dependent fluids such as generalised dark matter (GDM) scenarios \cite{Hu:1998kj,Kopp:2016mhm}. Our effective fluid can then describe a range of different DM particle models satisfying certain phase space relations \cite{Kunz:2016yqy,Kopp:2016mhm}. 
While the background entropy is constant for adiabatic fluids, spatial entropy perturbations represent genuine new primordial degrees of freedom specified through the initial conditions of the dark sector \cite{Kodama:1984ziu,Wands:2000dp,Malik:2004tf,Malik:2002jb,Celoria:2017xos}, affecting the subsequent cosmological dynamics. Nonetheless, when the coupling is switched off or the interaction rate is negligible, the entropy terms decouple from the dynamics and the matter fluid reduces to the standard CDM paradigm.

A key result of our analysis is that the background cosmological equations remain unchanged in the presence of entropy couplings, up to a trivial redefinition of the scalar field potential. 
This offers a consistent Lagrangian realisation of phenomenological models which modify $\Lambda$CDM only at the level of perturbations \cite{Richarte:2025tkj}.
Additionally, the continuity equation remains unmodified at linear order, while the coupling current vanishes non-linearly when projected along the dark matter fluid flow, implying no energy transfer in the fluid rest frame. Consequently, the entropy couplings modulate a pure-momentum exchange, similar to the Type-3 models of Ref.~\cite{Pourtsidou:2013nha} and velocity-entrainment models \cite{BeltranJimenez:2020qdu}, but arising from entropy gradients rather than velocity couplings.  
The leading modifications therefore manifest in the Euler equation, impacting the growth of matter perturbations. Indeed, we find that structure growth can be suppressed or enhanced on distinct scales, depending on the primordial scale dependence of the intrinsic entropy perturbations. This bears some resemblance to the phenomenological models with $k$-dependent dark matter modifications \cite{Arabameri:2023who}, except that in our case the scale dependence arises naturally from our dark sector interactions. The thermodynamic interpretation of the model is also appealing, giving rise to imperfect fluid properties at next-to-leading order in the effective energy-momentum tensor \cite{Misner:1973prb}. We discuss the physical and thermodynamic implications of these models throughout this work.

This paper is structured as follows. In \cref{sec:coup},  we first introduce the Brown fluid action and its modifications  within our non-minimally coupled Lagrangian framework, defining two classes of models: one with algebraic couplings and another with derivative couplings. We derive the corresponding equations of motion and coupling currents. In \cref{sec:cosmo}, we turn to cosmological applications, focusing on linear scalar perturbations. We present the general evolution equations for a perfect DM fluid with entropy couplings before specialising to the case of entropic-CDM. In \cref{sec:obs}, we compare the algebraic and derivative couplings and compute key observables, including the matter power spectrum, the CMB temperature spectrum, and the clustering amplitude $\sigma_8$. This allows us to discuss the distinct observational signatures associated with entropy interactions in the dark sector. Finally, in \cref{sec:disc}, we summarise our findings and outline future prospects.

\section{Entropy couplings and gravitational actions}
\label{sec:coup}

\subsection{Perfect fluid actions} 
The Brown Lagrangian formalism provides a covariant variational approach to describe relativistic perfect fluids within General Relativity \cite{Schutz:1977df,Brown:1992kc}.
This construction naturally yields the standard perfect fluid energy-momentum tensor and its corresponding equations of motion, with a consistent inclusion of additional thermodynamic degrees of freedom. 
The framework is therefore particularly useful for modelling the effective cosmological behaviour of dark matter and dark energy fluids, which may possess non-trivial internal structures or couplings at the microscopic level. 
In this section, we introduce the Brown Lagrangian formulation and comment on existing modifications in the literature, before presenting our new entropic couplings in \cref{sec:entropy_coupling}.

A cosmological perfect fluid, characterised by its particle number density $n$, energy density $\rho$, and entropy per particle $s$, can be described by the Brown action \cite{Brown:1992kc} 
\begin{equation} 
    S_{\textrm{Brown}} = \int \Big[-\sqrt{-g} \rho(n,s) + \mathcal{N}^\mu \left(\varphi_{,\mu} + s \theta_{,\mu} + \beta_{A}\alpha^{A}_{, \mu} \right) \Big]  \dd^4x \, , \label{S_Brown}
\end{equation}
where $\varphi$, $\theta$, and $\beta_{A}$ are Lagrange multipliers and
 $\alpha^{A}(x)$ serve as fluid coordinates to label the fluid flow lines at each spacetime point $x^{\mu}$. Subscript commas denote the partial derivative, e.g., $\varphi_{,} =\partial_{\mu} \varphi$. The densitised particle number flux $\mathcal{N}^{\mu}$ relates to the fluid four-velocity $u^{\mu}$ (satisfying $u^{\mu}u_{\mu}=-1$) and the particle number density $n$ via
\begin{align} \label{Js}
 \mathcal{N}^{\mu} = \sqrt{-g} n u^{\mu} \, , \qquad |\mathcal{N}| = \sqrt{-g_{\mu \nu} \mathcal{N}^{\mu} \mathcal{N}^{\nu}} \, , \qquad n= \frac{ |\mathcal{N}| }{\sqrt{-g}}\, ,
\end{align}
and is treated as a dynamical variable in place of $n$.
In the following section we will give the explicit variations with respect to all of the dynamical variables, but here we give a descriptive overview: the metric variations give rise to the standard perfect fluid energy-momentum tensor. The Lagrange multipliers $\varphi$ and $\theta$ enforce the conservation of particle number and entropy\footnote{Generalisations of the Brown action which violate number and entropy conservation have also been proposed, which can potentially describe a viable unified dark sector \cite{Iosifidis:2024ksa}.}, leading to the conservation laws $\nabla_{\mu}(n u^{\mu}) =0$ and $\nabla_{\mu}(s n u^{\mu})=0$. The remaining Lagrange multiplier $\beta_A$ restricts the fluid four-velocity to be directed along the hypersurface of constant $\alpha^{A}$. Finally, one can show that the variations with respect to the densitised particle number $\mathcal{N}^{\mu}$ and the entropy per particle $s$ are consistent with the covariant conservation of the energy-momentum tensor $\nabla^{\mu} T_{\mu \nu}=0$ \cite{Brown:1992kc}.

Modifications of the perfect fluid action have been widely used in the literature to model cosmological interactions, typically by introducing non-minimal couplings between fluid variables and external fields.
For instance, the \textit{algebraic} and \textit{derivative} coupled quintessence models were introduced by Boehmer et al. in Refs.~\cite{Boehmer:2015kta,Boehmer:2015sha} by supplementing the Brown action with a canonical scalar field Lagrangian and interaction terms of the form $\mathcal{L}_{\textrm{int}}(n,\phi,s)$ and $\mathcal{L}_{\textrm{int}}(n,\phi,s) \mathcal{N}^{\mu} \phi_{,\mu}$. These couplings were shown to modify the background cosmological evolution and give rise to phenomena such as screening mechanisms. 
Because these works focused only on the background cosmology with $s = \text{const.}$, entropy-related effects were not studied.
On the other hand, cosmological perturbations were studied for both classes of interacting model by Koivisto et al. in Ref.~\cite{Koivisto:2015qua}. However, the effects of entropy were neglected by assuming an isentropic constraint, instead of the more general adiabatic constraint implied by the fluid equations (see \cref{eq_s} and surrounding discussion). Many subsequent works have utilised these formalisms\footnote{Some works use the Schutz–Sorkin action formalism \cite{Schutz:1977df}, which is physically equivalent to the Brown action but is instead written using the velocity-potential representation. Both actions can be mapped to one another, see \cite{Brown:1992kc}.} to investigate the cosmological implications of non-minimally coupled interacting dark sectors \cite{Dutta:2017kch,Amendola:2020ldb,Kase:2019veo,Kase:2020hst,Liu:2023mwx,BeltranJimenez:2020qdu,Kase:2019mox}.

An alternative geometric approach used in the literature is the \textit{pull-back formalism} \cite{Comer:1993zfa,Comer:1994tw,Andersson:2020phh}, which can be reformulated in the modern language of effective field theory \cite{Dubovsky:2005xd,Dubovsky:2011sj,Ballesteros:2012kv}. In that case, the physical content remains the same, but one instead  works with variables defined in a three-dimensional fluid space (labelled by the fluid coordinates $\alpha^{A}(x)$ introduced above). 
Pourtsidou et al. \cite{Pourtsidou:2013nha} developed interacting dark sector models using this formalism by non-minimally coupling the number density $n$, scalar field $\phi$ and the two invariants $Y=\nabla_{\mu} \phi \nabla^{\mu}\phi$ and $Z=u^{\mu}\nabla_{\mu}\phi$. The latter invariant $Z$ is particularly interesting, as it couples the fluid four-velocity with a scalar field, leading to a type of momentum transfer. The Type-3 models of \cite{Pourtsidou:2013nha,Skordis:2015yra} take advantage of this coupling to study interacting models with a \textit{pure}-momentum transfer, which have the particularity of modifying the Euler equation for CDM but not the continuity equation (up to first order in cosmological perturbations).
Models with pure-momentum exchange, including the Type-3 models and the velocity-entrainment models\footnote{The models introduced in Ref.~\cite{BeltranJimenez:2020qdu} utilise a similar coupling to the Type-3 models, but instead couple the four-velocities of dark matter $u_{\mu}^{c}$ and dark energy $u_{\mu}^{d}$ via $Z=g_{\mu \nu} u_{\mu}^{c} u_{\mu}^{d}$. Such models also give rise a pure-momentum exchange \cite{BeltranJimenez:2021wbq}.} of Ref.~\cite{BeltranJimenez:2020qdu}, have seen extensive recent investigation in a range of cosmological contexts, both within and beyond the linear regime \cite{Linton:2017ged,Chamings:2019kcl,Linton:2021cgd,Tsedrik:2022cri,Sahoo:2025cvz,Luo:2025szq,Liu:2023rvo,BeltranJimenez:2021wbq,Kase:2019mox}. 

Most importantly, in all of the non-minimally coupled interacting frameworks studied so far, the effects of couplings mediated by entropy degrees of freedom have not been considered, despite its natural appearance in the Brown action \cref{S_Brown}. We therefore aim to extend these pioneering works by investigating couplings to the intrinsic entropy of dark matter, a framework that we present in the following sections.

\subsection{Entropic couplings and interactions} \label{sec:entropy_coupling}
In this work, we explore a new class of couplings that has not been previously studied in the literature, describing interactions between the intrinsic entropy per particle of dark matter, $s$, and an external scalar field $\phi$.
Our approach follows the previously discussed works of Boehmer et al. \cite{Boehmer:2015kta,Boehmer:2015sha} and Pourtsidou et al. \cite{Pourtsidou:2013nha}. We assume the perfect fluid to be non-isentropic $\partial_{\mu} s \neq 0$, and therefore our models are distinct from those previously studied. Instead, the intrinsic entropy per particle describes genuine degrees of freedom related to the internal structure of the dark matter fluid \cite{MUKHANOV1992203,Kodama:1984ziu,Malik:2004tf}. Consequently, its perturbations are also non-vanishing. A more thorough discussion on the interpretation of the intrinsic entropy and its perturbations is given in \cref{sec:entropy}. In this section, it suffices to treat the intrinsic entropy $s$ as an independent thermodynamic degree of freedom associated with the effective dark matter fluid.

Interactions between the intrinsic entropy and a scalar field can be written in the general form
\begin{equation} \label{general_L}
    \mathcal{L}_{\textrm{int}}(\phi,s,\nabla_{\mu} \phi, \nabla_{\mu}s,...) \, ,
\end{equation}
with the ellipses indicating higher order derivatives of $s$ and $\phi$. These can be thought of as `pure-entropy-scalar-field' interactions, with no other dynamical variables entering into the couplings.
Further extensions with mixed couplings of the form $\mathcal{L}(n,s,\phi,u^{\mu},...)$ are easily generalisable and would allow for more substantial modifications;
these would also impact the background equations \cite{Boehmer:2015kta,Boehmer:2015sha,Koivisto:2015qua} and would be interesting to investigate in relation to the cosmic tensions. In this work,  however, we focus exclusively on the effects of pure-entropy couplings, leaving these extended models to be investigated in the future. 

We focus on two particular classes of models within the general interacting entropy framework defined by \cref{general_L}: \textit{algebraic couplings} and linear \textit{derivative couplings}, given respectively by the following interaction Lagrangians
\begin{align}
 \label{S_int}
    \mathcal{L}_{\textrm{alg}} = g(s,\phi) \quad \text{and}\quad
    \mathcal{L}_{\textrm{deriv}} =  h (\nabla^{\mu} \phi \nabla_{\mu}s) \, ,
\end{align}
where $g$ and $h$ are arbitrary functions which specify the particular realisation of the model being considered. 
The algebraic models are clearly the simplest type of coupling that could be considered, while the derivative models have a similar form to the Type-3 momentum-coupling models of Ref.~\cite{Pourtsidou:2013nha}. What is remarkable is that the core property of the Type-3 models, whose coupling current modifies only the Euler equation to first-order in cosmological perturbations, is also shared by both our algebraic and derivative couplings. In fact, we will show that all entropy couplings of the form of \cref{S_int} leave the background dynamics completely unchanged, up to a redefinition of the scalar field's effective potential $V(\phi)$. Such models are phenomenologically appealing given the tight background cosmological constraints on energy exchange in the dark sector \cite{Valiviita:2009nu,Yang:2018xlt,Barros:2018efl}.
Our model therefore describes dark sector interactions that show up only at the level of linear cosmological perturbations.

\subsection{Field equations}
The total action is given by
\begin{equation}
    S_{\textrm{tot}} =  S_{\textrm{int}} + S_{\textrm{EH}} + S_{\phi} + S_{\textrm{Brown}}   + S_{\textrm{SM}}  \, ,
    \label{Stot}
\end{equation}
with the interaction Lagrangian 
\begin{equation}
     S_{\textrm{int}} = - \int  f(s,\phi,\mathcal{S}) \sqrt{-g}\, \dd^4x \, ,
\end{equation}
where $f$ includes both algebraic and derivative couplings through the scalar contraction $\mathcal{S} = \nabla^{\mu} \phi \nabla_{\mu} s$. 
The other terms in the total action are the Brown action of \cref{S_Brown}, an action for the Standard Model fields $S_{\rm SM}$, and the Einstein-Hilbert action and canonical scalar field action,
\begin{align}
S_{\textrm{EH}} &= \frac{1}{16 \pi G} \int \sqrt{-g}  R\, \dd^4x \, ,  \label{S_EH}  \\
        S_{\phi} &= -\int \sqrt{-g} \left( \frac{1}{2}\nabla_{\mu} \phi \nabla^{\mu} \phi + V(\phi) \right) \dd^4x \, . \label{S_phi}
\end{align}
In this section, we assume the Brown perfect fluid to describe dark matter, and neglect explicitly writing the other Standard Model matter components. 
In \cref{sec:cosmo} we will reintroduce the (uncoupled) matter components for the baryonic and relativistic particle species, and will then include labels on the matter variables to denote each specific component. 

The variations of the total action in \cref{Stot} with respect to the independent variables $g_{\mu \nu}$, $\mathcal{N}^{\mu}$, $s$, $\varphi$, $\theta$, $\alpha^A$, $\beta_{A}$, and $\phi$ are
\begin{align} \label{deltag}
    \delta g^{\mu \nu} \quad &: \quad \frac{1}{16 \pi G} G_{\mu \nu} - \frac{1}{2}T_{\mu \nu} - \frac{1}{2}  T_{\mu \nu}^{(\phi)} -  \frac{1}{2} T_{\mu \nu}^{(\textrm{int})} = 0 \, , \\ \label{deltaJ}
    \delta \mathcal{N}^{\mu} \quad &: \quad  \mu u_{\mu} + \varphi_{, \mu} + s \theta_{,\mu} + \beta_{A} \alpha^{A}_{, \mu} =0 \, , \\
    \delta s \quad &: \quad n u^{\mu} \theta_{, \mu} -  \frac{\partial \rho}{\partial s} + \nabla_{\mu}\left(\frac{\partial f}{\partial \mathcal{S}} \nabla^{\mu} \phi \right) - \frac{\partial f}{\partial s} =0 \, , \label{deltas} \\
     \delta \varphi \quad &: \quad -\mathcal{N}^{\mu}_{, \mu}  =0 \, , \label{deltaphi} \\
     \delta \theta \quad &: \quad -(s \mathcal{N}^{\mu})_{, \mu} =0 \, , \label{deltatheta} \\
      \delta \beta_{A} \quad &: \quad  \alpha^{A}_{,\mu}\mathcal{N}^{\mu} =0 \, , \label{deltaalpha} \\
    \delta \alpha^{A} \quad &: \quad - (\beta_{A}\mathcal{N}^{\mu})_{,\mu} =0 \, , \label{deltaBeta} \\
     \delta \phi \quad &: \quad \Box \phi - V'(\phi) +\nabla_{\mu}\left( 
     \frac{\partial f}{\partial S}\nabla^{\mu}s\right) - \frac{\partial f}{\partial \phi} =0 \, , \label{phiVar}
\end{align}
where $T_{\mu \nu}$ and $ T_{\mu \nu}^{(\phi)}$ are the usual perfect fluid and scalar energy-momentum tensors, and $ T_{\mu \nu}^{(\textrm{int})}$ is the new interaction tensor
\begin{align} \label{Tperf}
    T_{\mu \nu} &= \left(\rho + p \right) u_{\mu} u_{\nu} +  g_{\mu \nu} p \, , \\ \label{Tscalar}
    T_{\mu \nu}^{(\phi)} &= \nabla_{\mu} \phi \nabla_{\nu} \phi - \frac{1}{2} g_{\mu \nu} \nabla_{\lambda}\phi \nabla^{\lambda}\phi - g_{\mu \nu} V(\phi) \, ,  \\ 
    T_{\mu \nu}^{(\textrm{int})} &= - g_{\mu \nu} f + 2 \frac{\partial f}{\partial \mathcal{S}} \nabla_{(\mu} \phi \nabla_{\nu)} s  \, . \label{Tint}
\end{align}
In the above, we have also made use of the following thermodynamic definitions for the matter fluid
\begin{align} \label{pressure}
    \textrm{pressure:} & \ p = n \frac{\partial \rho}{\partial n} - \rho \, , \\
    \textrm{chemical potential:} & \ \mu = \frac{\rho + p}{n} \, .
\end{align}
The Lagrange multipliers $\varphi$ and $\theta$ enforce the conservation of particle number flux and entropy along the fluid flow lines $u^{\mu} = \mathcal{N}^{\mu}/(n \sqrt{-g})$ respectively
\begin{equation} \label{cons}
    \nabla_{\mu}(n u^{\mu}) =0 \, , \qquad \nabla_{\mu}(n  s u^{\mu}) =0 \, .
\end{equation}
Taken together, this implies that the entropy per particle is constant when \textit{projected along the fluid flow}
\begin{equation} \label{entropy}
    u^{\mu} \nabla_{\mu} s = 0 \, .
\end{equation}
As discussed in \cref{sec:entropy}, this does not generally imply that the entropy is constant or non-dynamical. On cosmological backgrounds one can straightforwardly see that \cref{entropy} implies $s = \textrm{const.}$, but this is not true at the level of perturbations. In contrast to isentropic fluids, the linear perturbations of adiabatic fluids can be spatially varying $\delta s = \delta s(\vec{x})$. This will be shown explicitly in the next section, where we investigate the consequences of non-vanishing intrinsic entropy perturbations. As such, $\delta s$ is required to have its own independent initial conditions, analogous to the more well-known isocurvature modes \cite{Malik:2002jb,Malik:2004tf,Kodama:1984ziu}. In past works on perturbations in non-minimally coupled models such as Ref.~\cite{Koivisto:2015qua}, the stricter isentropic constraint $\partial_{\mu} s=0$ has been assumed. Therefore, our results extend those found in the current literature.  

Comments on the physical interpretation of interactions can now be made. From \cref{Tint} we can define the energy density, isotropic pressure, heat flux and anisotropic stress of the interaction energy-momentum tensor \cite{Misner:1973prb}. These next-to-leading order (NLO) terms can then be connected with phase space distributions \cite{Strickland:2014pga,Kopp:2016mhm}. We employ the standard covariant $1+3$ decomposition with respect to the DM fluid four-velocity $u^{\mu}$ along with the  spatial metric (projection operator) $h_{\mu \nu}  = g_{\mu \nu} + u_{\mu} u_{\nu}$
to write \cite{Ellis:1998ct}
\begin{equation}
     T_{\mu \nu}^{(\textrm{int})} = \rho^{(\textrm{int})} u_{\mu} u_{\nu} + h_{\mu\nu} p^{(\textrm{int})} + 2 u_{(\mu} q_{\nu)}^{(\textrm{int})} + \pi_{\mu\nu}^{(\textrm{int})} \, ,
\end{equation}
where $q_{\mu}^{(\textrm{int})}$ and $\pi_{\mu\nu}^{(\textrm{int})}$ are the heat flux and anisotropic stress, respectively.
The interaction energy density and isotropic pressure are
\begin{align} \label{int_rho}
    \rho^{(\textrm{int})} &:= u^{\mu} u^{\nu}  T_{\mu \nu}^{(\textrm{int})} = f \, , \\
  p^{(\textrm{int})} &:= \frac{1}{3} h^{\mu \nu}  T_{\mu \nu}^{(\textrm{int})} = -f + \frac{2}{3} f_{,\mathcal{S}} \nabla_{\mu}\phi \nabla^{\mu}s \, . \label{int_p}
\end{align}
Note that we have used the adiabatic constraint, \cref{entropy}, which implies $h^{\mu \nu} \nabla_{\nu} s = \nabla^{\mu}s$
to simplify the results above. 
The heat flux and anisotropic stress are
\begin{align} \label{int_q}
     q_{\mu}^{(\textrm{int})} &:= u^{\nu} h^{\lambda}_{\mu} T_{\nu\lambda}^{(\textrm{int})} = f_{,\mathcal{S} }  u^{\nu} \nabla_{\nu} \phi  \nabla_{\mu} s \, , \\
     \pi_{\mu\nu}^{(\textrm{int})} &:= T_{\langle \mu \nu \rangle}^{(\textrm{int})} = 2 f_{,\mathcal{S}} \nabla_{\langle \mu} \phi \nabla_{\nu \rangle} s \label{int_pi} \, ,
\end{align}
where angled brackets denote the orthogonally projected, symmetric, trace-free part of an arbitrary rank-two tensor $A^{\mu \nu}$ \cite{Ellis:1998ct}
\begin{equation}
A^{\langle \mu \nu \rangle} := \big( h^{(\mu}_{\lambda}h^{\nu)}_{\rho}- \frac{1}{3} h^{\mu \nu} h_{\lambda \rho} \big) A^{\lambda \rho} \, ,
\end{equation}
which picks out only spatial components.
The heat flux and anisotropic stress will be first- and second-order terms in cosmological perturbations respectively, since the spatial derivatives of $\phi$ and $s$ are both first-order quantities. In the isentropic limit, all of the above terms either vanish or simply reduce to functions of $\phi$, which can then be absorbed into the scalar field potential in \cref{Tscalar}.
 
It is illustrative to compare the results of \cref{int_rho,int_p,int_q,int_pi} with the similar Type-3 models, which have a modified effective density and pressure (Eqs. (73) and (54) of \cite{Pourtsidou:2013nha}) and non-vanishing heat flux and anisotropic stress. Both models give rise to an effective density and isotropic pressure at leading order, non-vanishing heat fluxes at linear order, and anisotropic stresses at second-order. However, the background dynamics are modified in the Type-3 models, differing from our setup. Similarly, at the linear level, the source of interactions in the Type-3 models are driven by spatial gradients of the scalar fields; ours are instead driven by the spatial gradients of the intrinsic entropy per particle, which has a distinctly different origin and will lead to different cosmological dynamics.

Given these non-vanishing NLO terms in \cref{int_q,int_pi}, it is also worthwhile to compare our results to other studies going beyond the perfect fluid approximation, e.g., cosmological models with imperfect dark sector fluids \cite{Koivisto:2005mm,Li:2009mf}. The modified cosmological dynamics that we study in \cref{sec:cosmo} will share some resemblance in structure with these models.
Similarly, if we view these modifications as a change in the effective dark matter fluid description, one can make direct comparisons with the Generalised Dark Matter models \cite{Hu:1998kj,Kopp:2016mhm}.
It is interesting, however, that our modifications originate from a simple coupling between the perfect fluid DM and scalar field sectors.
For more details on relativistic fluids, including the many connections that can be made between heat fluxes, viscosity, and entropy, see \cite{Rezzolla:2013dea,Bemfica:2020zjp,Blas:2015tla,Pimentel:2016jlm,Bhattacharya:2012zx}. 
For the purpose of this work, we will treat the modifications as part of the scalar field sector\footnote{One can similarly re-derive \cref{int_rho,int_p,int_q,int_pi} in the scalar field frame with four-velocity $u^{(\phi)}_{\mu}$ \cite{Faraoni:2012hn,Faraoni:2022gry}, giving non-vanishing contributions to the effective NLO terms.}.

The full metric field equations, following \cref{deltag},
are 
\begin{align}
    G_{\mu \nu} &= 8 \pi G \big(T_{\mu \nu} + T_{\mu \nu}^{(\phi)} + T_{\mu \nu}^{(\textrm{int})} \big) \, , 
\end{align}
and the scalar field equation, \cref{phiVar}, is
\begin{equation}  \label{scalarEoM}
     \Box \phi - V'(\phi) +\nabla_{\mu}\left( f_{,\mathcal{S} }
     \nabla^{\mu}s\right) - f_{,\phi }  = 0 \, .
\end{equation}
The dark matter and interaction tensors are implicitly coupled through the dependence of the density $\rho$ and pressure $p$ on the entropy $s$, whereas the coupling between $s$ and $\phi$ is explicit in \cref{Tint}. 
Let us define the total modified scalar-field energy-momentum tensor as 
\begin{equation} \label{TDE}
    \tilde{T}^{(\phi)}_{\mu \nu}:= T_{\mu \nu}^{(\phi)} + T_{\mu \nu}^{(\textrm{int})}  \, ,
\end{equation}
which we take to represent the effective dark energy fluid. An important caveat is that the split between dark matter and dark energy is essentially arbitrary, and so the presence of energy or momentum exchange is a matter of definitions \cite{Pourtsidou:2013nha,Boehmer:2015kta,Tamanini:2015iia}. We choose to group terms with the dark matter number density $n$ in the matter sector, and those with a scalar field $\phi$ into the dark energy sector. Pure-entropy couplings therefore do not modify the definition of the dark matter fluid.
It is in this way that our results differ from the usual energy-exchange predicted from non-minimally coupled models with number density couplings, such as \cite{Boehmer:2015kta,Boehmer:2015sha,Koivisto:2015qua}.

From the contracted Bianchi identity it follows that $\nabla^{\mu} \tilde{T}_{\mu \nu}^{(\phi)} = -\nabla^{\mu} T_{\mu \nu}$, and the interaction current is thus defined as
\begin{equation}
    J_{\nu} := \nabla^{\mu} \tilde{T}_{\mu \nu}^{(\phi)} = -\nabla^{\mu} T_{\mu \nu} \, .
\end{equation}
The detailed calculations of these conservation equations and the proof that they follow exactly from the variational equations, \cref{deltag,deltaJ,deltas,deltaphi,deltatheta,deltaalpha,deltaBeta,phiVar}, is given in \cref{append_cons}. 
Note that we will also have the uncoupled matter tensor being conserved along the fluid flow as usual
\begin{equation}
    u^{\mu} \nabla^{\nu} T_{\mu \nu} = 0 \, ,
\end{equation}
which follows directly from the conservation of number flux and entropy, \cref{cons}. This implies that the projection along the fluid velocity of the coupling current should also vanish $u^{\mu} J_{\mu} =0$, which we confirm below and show explicitly in \cref{append_cons} directly from the variational equations. This result can also be found in Ref.~\cite{Boehmer:2015kta}.

The interacting coupling current $J_{\mu}$ is calculated by taking the covariant derivative of the total (modified) scalar field energy-momentum tensor and making use of the scalar field equations of motion in \cref{scalarEoM}. For the (uncoupled) scalar field energy-momentum tensor we easily obtain
\begin{equation} \label{Jphi}
     \nabla^{\mu}T_{\mu \nu}^{(\phi)}= \left( f_{,\phi} -\nabla_{\mu}( f_{,\mathcal{S} }  \nabla^{\mu}s) \right)  \nabla_{\nu} \phi \, .
\end{equation}
For the interaction term we have 
\begin{align}
\nabla^{\mu}T_{\mu \nu}^{(\textrm{int})} &= \nabla_{\mu}\big(f_{,\mathcal{S} }  \nabla^{\mu}s) \nabla_{\nu} \phi + \nabla_{\mu}\big(f_{,\mathcal{S} }   \nabla^{\mu} \phi \big) \nabla_{\nu} s - f_{,\phi} \nabla_{\nu} \phi  -f_{,s} \nabla_{\nu} s \, ,
\end{align}
with the first and third terms exactly cancelling the contribution from \cref{Jphi}. 
Our total modified scalar field conservation equation and the coupling current is therefore
\begin{align}\label{Jcouple}
   J_{\nu} &:=  \nabla^{\mu} \tilde{T}_{\mu \nu}^{(\phi)}  = \nabla_{\mu}\big(f_{,\mathcal{S} }  \nabla^{\mu} \phi \big) \nabla_{\nu} s - f_{,s} \nabla_{\nu} s
    \, .
\end{align}
One immediately verifies that its projection along the fluid velocity vanishes $u^{\nu} J_{\nu}=0$ due to $u^{\nu} \nabla_{\nu} s = 0$. We therefore have a current that only acts in the perpendicular direction
\begin{equation} \label{JcoupPerp}
    J_{\nu} = h_{\nu}{}^{\mu} J_{\mu} =\big(
    \nabla_{\lambda}(f_{,\mathcal{S} }  \nabla^{\lambda} \phi)  - f_{,s} \big)  h_{\nu}{}^{\mu} \nabla_{\mu} s
    \, .
\end{equation}
These relations satisfy total conservation of energy-momentum, which is enforced both by the field equations with the contracted Bianchi identity and from the fluid variable variations in \cref{deltaJ,deltas,deltaphi,deltatheta,deltaalpha,deltaBeta} (see \cref{append_cons}). Our results agree with the general conclusions that the uncoupled energy-momentum tensor in interacting Brown fluid models is conserved along the fluid flow \cite{Boehmer:2015kta,Boehmer:2015sha}.
A similar equation appeared recently in the $f(R,T)$ modified gravity models, which gives a purely-perpendicular current sourced by entropy couplings \cite{Boehmer:2025afy}. In summary, pure-entropy couplings generate trivial vacuum-like background contributions, pure-momentum exchange at linear order, and a conserved coupling current along the DM-fluid flow at all orders.

Lastly, it is important to note that the coupling current and dark sector interactions are switched off in two cases. The first is when the interaction term vanishes or does not depend on $s$, which is trivial to see. The second case is when the dark matter fluid energy density does not depend on the entropy $\rho=\rho(n)$. Consequently, the entropy variation \cref{deltas} will not include a  $\partial \rho/\partial s$ term, and therefore the coupling term will not enter into the covariant conservation of the energy-momentum tensor. This is clear from \cref{append_rhos} in \cref{append_cons}, where it is shown explicitly that the perpendicular projection of the matter divergence is proportional to $\partial \rho/\partial s \nabla_{\nu} s$. In what follows, we therefore always assume that the dark matter energy density depends on the intrinsic entropy $\rho=\rho(n,s)$. For numerical implementations, we specialise to a particular dark matter equation of state that depends on the entropy, which we call \textit{entropic-CDM}, whose properties are discussed in \cref{sec:entropy} and summarised in \cref{table}.

\section{Cosmology} \label{sec:cosmo}

We now turn to cosmological applications. Let us choose the interaction to couple the dark energy scalar field with the dark matter intrinsic entropy $s$ through $f(\phi, s, \nabla_{\mu} \phi \nabla^{\mu}s)$. We first study the background cosmological equations before examining scalar perturbations. Note that we assume all other matter components (i.e., baryons $\rho_b$ and relativistic species $\rho_r$, including radiation and neutrinos) to be uncoupled as usual. We will generically refer to the dark matter fluid as $\rho_c$, with number density and entropy labelled as $n$ and $s$ respectively (with no subscript), as these thermodynamic variables will not be relevant for the description of the other species. In the later sections, we will specify our dark matter model to be described by the entropic-CDM fluid with barotropic equation of state, see \cref{sec:entropy}. However, we first present the completely general analysis, applicable to any perfect fluid dark matter model, which may possess non-vanishing sound speeds or non-adiabatic pressure perturbations.

\subsection{Background}
We first look at the background cosmology, where the effects of entropy couplings are trivial and the dynamics are unchanged from the uncoupled quintessence scenario. In this section, all variables are taken to be evaluated on the FLRW background 
\begin{equation}
    \dd s^2 = a(\tau)^2 \left[ -\dd \tau^2 +  \dd x^2 + \dd y^2 + \dd z^2 \right] \, ,
\end{equation} 
where $a(\tau)$ is the scale factor.
For adiabatic fluids satisfying \cref{entropy}, the background entropy is constant $\nabla_{\mu} s =0$. It follows that the coupling current in \cref{Jcouple} also vanishes at the background $J_{\mu} =0$. Similarly, because $\mathcal{S}=\nabla^{\mu}\phi \nabla_{\mu}s = 0$ at the background, all contributions from the interaction energy-momentum tensor, \cref{Tint}, amount to a simple redefinition of the effective scalar field potential. The evolution of background quantities is therefore unchanged from the standard quintessence scenario. This property holds independently of the functional form of the coupling and is generic to the family of entropy couplings considered in this work. This implies that entropy couplings cannot be traced solely by background probes and the constraints on the model's parameters will necessarily come from the evolution of observables related to the perturbations.

Firstly, we define the conformal Hubble parameter as
\begin{equation} \label{hubble}
    \mathcal{H}(\tau) = \frac{a'}{a} \, ,
\end{equation}
where a prime denotes the derivative with respect to conformal time $\tau$.
The modified energy-momentum tensor for dark energy, \cref{TDE}, can be used to define the background energy density and pressure. These take the standard quintessence form
\begin{align}
    \rho_{\textrm{DE}} =  \frac{1}{2a^2} {\phi}'^2 + \hat{V}(\phi)\, , \qquad 
    p_{\textrm{DE}} = \frac{1}{2 a^2} {\phi}'^2 - \hat{V}(\phi) \, ,
\end{align}
where we have redefined the effective scalar potential $\hat{V}(\phi) = V(\phi) + f(\phi, s, 0)$ in terms of the constant background dark matter intrinsic entropy $s$. 
The background scalar field equation, \cref{phiVar}, is 
\begin{equation} \label{phi_back}
    {\phi}'' + 2 \mathcal{H} {\phi}' + a^2 \hat{V}_{,\phi} = 0 \, .
\end{equation}
which can be written as $\Box \phi = \hat{V}_{,\phi}$. Similarly, the background Einstein field equations take the form
\begin{align}
    3H^2  &= 8\pi G  a^2 \Big( \sum_{c,b,r} \rho_{I} + \rho_{\textrm{DE}} \Big) \, ,
    \label{EFE1b} \\
    2 \mathcal{H}' + \mathcal{H}^2 &= -8\pi G  a^2 \Big( \sum_{c,b,r} p_{I}  + p_{\textrm{DE}}\Big) \, , \label{EFE2b}
\end{align}
where the index $I$ sums over the standard cosmological fluid components, comprised of dark matter $\rho_{c}$ and the
Standard Model fields (baryons $\rho_{b}$ and relativistic species $\rho_{r}$).
 The background continuity equation follows from $J_{\mu} =0$, or equivalently from the field equations, \cref{EFE1b,EFE2b}, and takes its standard form
\begin{equation} \label{contback}
    {\rho_{I}}' + 3 \mathcal{H} (\rho_{I}+p_{I}) =0 \, .
\end{equation}
This holds for each matter fluid component $I=c,b,r$, as well as the effective dark energy fluid, reflecting the absence of couplings at the background level.

\subsection{Scalar perturbations}
\subsubsection{Definitions}
Let us now study scalar perturbations, working in the conformal Newtonian gauge
\begin{equation}
    \dd s^2 = a(\tau)^2 \left[ - (1 + 2 \Phi) \dd \tau^2 +  (1-2\Psi) (\dd x^2 + \dd y^2 + \dd z^2) \right] \, ,
\end{equation}
with the scalar modes $\Phi$ and $\Psi$ depending on all coordinates. The corresponding expressions in the synchronous gauge are presented in \cref{sec:sync}.
We perturb the dark matter fluid variables and the scalar field to linear order
\begin{equation} \label{eq_pert}
    \phi(\tau,\vec{x}) = \phi(\tau) + \delta \phi(\tau,\vec{x}) \, , \quad n(\tau,\vec{x}) = n(\tau) + \delta n(\tau,\vec{x}) \, , \quad s(\tau,\vec{x}) = s(\tau) + \delta s(\tau,\vec{x}) \, , 
\end{equation}
with the homogeneous background quantities only depending on time and the perturbed quantities depending on all coordinates. 
Perturbations of the energy density $\rho(\tau,\vec{x}) = \rho(\tau) + \delta \rho(\tau,\vec{x})$ and pressure $p(\tau,\vec{x}) = p(\tau) + \delta p(\tau,\vec{x})$ are then implicitly defined from \cref{eq_pert}, see \cite{MUKHANOV1992203,Kodama:1984ziu}. Linear perturbations of the matter four-velocity $u_{\mu}(\tau,\vec{x}) = u_{\mu}(\tau) + \delta u_{\mu}(\tau,\vec{x})$ have components
\begin{equation}
    u_{\mu} = (-a,0,0,0) \, , \quad \delta u_{\mu} = a (-\Phi, v_{i}  ) \, ,
\end{equation}
where the coordinates dependence will be dropped from quantities where clear. 

The perturbed matter energy-momentum tensor for each component up to linear order is
\begin{equation}
\begin{aligned}
    \delta T^{0}_{I}{}_{0} &= -\delta \rho_{I} \, , \quad & \delta T^{i}_{I}{}_{0} &= -v^i_I (\rho_I+p_I) \, , \\ 
    \delta T^{0}_{I}{}_{i} &=  v^i_I (\rho_{I}+p_{I})  \, , \quad & \delta T^{i}_{I}{}_{j} &=  \delta p_{I} \delta^{i}_{j} +\Sigma^{i}_{I \, j}\, ,
\end{aligned}
\end{equation}
with traceless anisotropic stress satisfying $\Sigma^{i}_{I \,i}=0$.
For dark matter or baryons after decoupling, the fluid is perfect with vanishing shear $\Sigma^{i}_{I \,j}=0$. In these cases, the perturbed matter tensor can be directly obtained by perturbing the fundamental fluid variables $n$ and $s$ and performing simple algebraic manipulations \cite{Kodama:1984ziu}. For other imperfect cosmological species, such as neutrinos or photons after decoupling, the anisotropic stress is non-vanishing \cite{Ma:1995ey}.
Lastly, let us also define the fractional density contrast and velocity divergence of each fluid component
\begin{equation}
    \delta_{I} := \frac{\delta \rho_{I}}{\rho_{I}} \, , \qquad \theta_{I} := \partial_{i} v^{i}_{I} \, .
\end{equation}
When going to Fourier space, we use the convention $\nabla^2 = \partial_i \partial^i =  -k^2$.

\subsubsection{Particle number and entropy conservation}

Let us now examine the cosmological particle and entropy number conservation equations, which are equivalent with energy conservation. These relations hold for each individual matter fluid component that can be modelled as a perfect fluid, but we focus on the results explicitly for dark matter with $n$ and $s$.
Beginning with the number density equation, \cref{cons}, up to linear order in perturbations we have 
\begin{equation} \label{ncons}
     a^3 \Big( 3 \mathcal{H} n + n' + 3 \mathcal{H} (\delta n - 3 n \Psi ) - 3 \Psi n' + n \partial_{i} v^{i}_{c} + {\delta n}' - 3 n {\Psi}'\Big) = 0  \, ,
\end{equation}
where the background part immediately gives the usual scaling $n \propto a^{-3}$.
The entropy conservation equation, \cref{entropy}, to first order is
\begin{equation} \label{scons}
    \frac{1}{a}\big( s' - s' \Phi +\delta s' \big) = 0 \, ,
\end{equation} 
which shows that the background entropy must be constant $s = \text{const.}$ and the entropy perturbations can depend only on the spatial coordinates $\delta s = \delta s(\vec{x})$. Note that $s'=0$ implies that the entropy perturbation $\delta s$ is automatically gauge-invariant \cite{Bean:2003fb,kambe2007gauge}. The entropy perturbations are allowed to have spatially varying initial conditions, which do not evolve in time and can be interpreted as remnants of the fluid's thermodynamic properties. For instance, these could be set by early-universe processes, with non-zero intrinsic entropy emerging from quantum fluctuations during inflation \cite{Cicoli:2021itv} or phase transitions \cite{Lyth:2009imm}.
In fact, non-zero intrinsic entropy perturbations could be associated with all of the matter components, but in general the dynamics associated with intrinsic entropy perturbations completely decouple for adiabatic fluids; the key exception is when the entropy is non-minimally coupled explicitly to external fields. This is why the entropy contributions of the uncoupled adiabatic fluids $s_{b}$ and $s_r$ can be neglected from further discussion. On the other hand, for the non-minimally coupled dark matter entropy $s$, these contributions will affect the overall cosmological dynamics.

The contraction of the scalar field and entropy gradients $\mathcal{S}(\tau,\vec{x}) = \nabla^{\mu}\phi(\tau,\vec{x})  \nabla_{\mu}s(\tau,\vec{x})$ vanishes at linear order
\begin{align} \label{S_lin}
    \delta \mathcal{S} = -\frac{{s'}{\phi}'}{a^2} + \frac{1}{a^2} \Big( \phi' (2 \Phi s' - {\delta s}' ) - {s}' {\delta \phi}' \Big) = 0 \, ,
\end{align} 
following straightforwardly from the entropy conservation equation, \cref{scons}.
On the other hand, it does not vanish at second order in cosmological perturbations $\delta \mathcal{S}^{(2)} \neq 0$. For the purely-derivative models $f=f(\mathcal{S})$, we are therefore required to choose Lagrangians such that $f'(0) \neq 0$ such that the interaction tensor \cref{Tint} is non-vanishing at first-order.

\subsubsection{Dark energy perturbations}

For the linear perturbations of the effective dark energy fluid, \cref{TDE}, it is first convenient to explicitly split the model into its algebraic and derivative parts
\begin{equation}
    f(\phi, s, \nabla_{\mu} \phi \nabla^{\mu}s) = g(\phi, s) + h(\nabla_{\mu} \phi \nabla^{\mu}s) \, .
    \label{eq:f_function}
\end{equation}
In the linear cosmological equations, the functions $g$ and $h$ will depend only on background variables $\phi(\tau)$ and $s=\text{const.}$, since they will always be multiplied by first-order quantities $\delta \phi$ or $\delta s$. It follows that $g$ and $h$ will depend only on time $\tau$ through their dependence on the scalar field.
Using \cref{eq:f_function}, the density and pressure perturbations are be given by
\begin{align} 
    \delta \rho_{\textrm{DE}} &= \frac{{\phi'} \big( {\delta \phi}' - \phi' \Phi \Big)}{a^2}  + \delta \phi \hat{V}_{,\phi}   + g_{,s} \delta s  \, , \label{rho_DE} \\  \label{p_DE}
    \delta p_{\textrm{DE}} &= \frac{{\phi'} \big( {\delta \phi}' - \phi' \Phi \Big)}{a^2} - \delta \phi \hat{V}_{,\phi} -   g_{,s}\delta s \, ,
\end{align}
which are modified only by the algebraic couplings $ g(\phi,s)$.
The dark energy density contrast and velocity are
\begin{align} 
    \delta_{\textrm{DE}} &= 2\frac{a^2 \Big( \delta \phi  \hat{V}_{,\phi}  + g_{,s} \delta s \Big) + {\phi'} \left( {\delta \phi}'   - \phi' \Phi \right)}{2a^2 \hat{V}(\phi) + \phi'^2 } \, , \label{delta_DE}  \\
    v^{i}_{\textrm{DE}} &= - \frac{\partial^{i} \big(h_0 \delta s  + \delta \phi \big)}{\phi'} \, , \label{v_DE} 
\end{align}
where we define the constant $h_0 = h'(\mathcal{S})$. The derivative couplings can therefore only enter as a constant parameter $h_0$ due to $\delta\mathcal{S}$ vanishing at linear order in \cref{S_lin}.
Generically, we see that algebraic couplings modify the energy density and pressure, while derivative couplings enter through the momentum sector. Note that the shear contribution vanishes $\Sigma^{i}_{\textrm{DE}\, j}=0$, following from \cref{int_pi}.

While the background equation of state $w_{\textrm{DE}}= p_{\textrm{DE}}/ \rho_{\textrm{DE}}$ and adiabatic sound speed $c_{a,\textrm{DE}}^2 = p_{\textrm{DE}}' / \rho_{\textrm{DE}}'$ are unchanged from standard quintessence,
\begin{equation}
    w_{\textrm{DE}}  = \frac{\phi'^2 - 2a^2 \hat{V}}{\phi'^2 + 2a^2 \hat{V}} \, ,  \quad c_{a,\textrm{DE}}^2  = w- \frac{w'}{3\mathcal{H}(1+w)} \, , 
\end{equation}
the effective sound speed of the dark energy fluid is modified by the entropy couplings (see \cref{c_s}). This can be easily seen by rewriting the pressure perturbations as
\begin{equation}
    \delta p_{\textrm{DE}} = \delta \rho_{\textrm{DE}} - 2 \delta \phi \hat{V}_{,\phi} - 2 g_{,s} \delta s \, ,
\end{equation}
from which we can immediately identify new terms related to non-adiabatic pressure perturbations  \cite{Kodama:1984ziu}. The dark energy density and pressure perturbations in the rest frame 
(comoving gauge defined by $\tilde{T}^{(\phi)0}{}_{i}|_{\textrm{(rf)}} = 0$  \cite{Valiviita:2008iv}) are related to an arbitrary gauge by the transformations 
\cite{Kodama:1984ziu,Bean:2003fb}
\begin{align} \label{rest_trans}
    \delta \rho_{\textrm{DE}}|_{\textrm{(rf)}} =  \delta \rho_{\textrm{DE}} - \rho_{\textrm{DE}}' \frac{\theta_{\textrm{DE}}}{k^2} \, , \quad \delta p_{\textrm{DE}}|_{\textrm{(rf)}} =  \delta p_{\textrm{DE}} - p_{\textrm{DE}}' \frac{\theta_{\textrm{DE}}}{k^2}   \, ,
\end{align}
with $\theta_{\textrm{DE}}= \partial_{i}v^{i}_{\textrm{DE}}$ computed from \cref{v_DE}.
Note that if the background was coupled, additional terms would enter via $ \rho_{\textrm{DE}}'$ and $p_{\textrm{DE}}'$ \cite{Valiviita:2008iv}.
We can then straightforwardly calculate the gauge-invariant rest frame sound speed 
\begin{align} \label{cs_de}
    \hat{c}_{s,\textrm{DE}}^2 := \frac{  \delta p_{\textrm{DE}}}{\delta \rho_{\textrm{DE}}} \Bigg|_{\textrm{(rf)}} 
    = 1 - \frac{2 \delta s (g_{,s} - h_0 \hat{V}_{,\phi})}{\delta \rho_{\textrm{DE}}|_{\textrm{(rf)}} }
    &=  1 - \frac{2 a^2 k^2 \delta s (g_{,s} - h_0 \hat{V}_{,\phi})}{a^2 k^2 \delta \rho_{\textrm{DE}} + 3 \mathcal{H} \phi'^2 \theta_{\textrm{DE}}} \, , 
\end{align}
where we have used the background continuity equation for dark energy, \cref{contback}.
The final equality is written explicitly in terms of our conformal Newtonian gauge variables. 
In the absence of entropic couplings $g_{,s}=h_0=0$, the effective sound speed squared takes its usual value $\hat{c}_{s,\textrm{DE}}^2= 1$. However, in general, both derivative and algebraic couplings modify $\hat{c}_{s,\textrm{DE}}^2$. We will shortly show that this combination of $g_{,s}$ and $h_0$ appearing in the numerator of \cref{cs_de} can only vanish when the coupling current vanishes, reducing to the trivial uncoupled case. This differs from some of the Type-3 pure-momentum coupling models, which can leave the sound speed unchanged (e.g., for the quadratic models \cite{Linton:2017ged}).

Lastly, it should be reminded that $\hat{c}_{s,\textrm{DE}}^2$ is not directly related to stability because the dark energy fluid has a non-adiabatic component and hence a different adiabatic sound speed \cite{Valiviita:2008iv}.
The physical propagation speed is determined by the kinetic structure of the model, which requires studying the quadratic action. As discussed in \cite{DeFelice:2009bx,DeFelice:2010sh}, the appropriate perturbation variables in the reduced quadratic action are generically not the number density and entropy per particle; in our Brown action \cref{S_Brown}, for instance, one instead would use the Lagrange multipliers $\varphi$ and $\theta$. The non-adiabatic degree of freedom in the kinetic matrix is therefore not expected to be $\delta s$, so the vanishing of $\delta s'$ does not immediately imply a strong coupling problem. For algebraic entropy couplings, no new derivative interactions are introduced, while for derivative couplings the kinetic structure requires a more careful treatment. More generally, the no-ghost and gradient stability conditions will imply constraints on the underlying fluid model $\rho(n,s)$ and the couplings $g$ and $h$. A full derivation of the quadratic action is beyond the scope of the present work and will be investigated in detail in a forthcoming publication. For the purposes of the observational analysis in \cref{sec:obs}, we numerically verify that dark matter and dark energy perturbations remain finite and do not exhibit rapid runaway growth over the scales of interest.

\subsection{Equations of motion}
\subsubsection{Coupling current and evolution equations}

Let us now examine the coupling current $J_{\mu}$ of \cref{JcoupPerp} for cosmological perturbations. It is straightforward to calculate this to linear order in cosmological perturbations 
\begin{align} \label{Jpert}
     \delta J_{0} = 0 \, , \qquad 
    \delta J_{i} = \left(h_0 \hat{V}_{,\phi}- g_{,s}  \right)  \partial_{i} \delta s
    \, ,
  \end{align}
where we have used the entropy equation, \cref{scons}, to eliminate $s'$ and $\delta s '$. We therefore see that the coupling current has non-vanishing contributions up to first order in the spatial perturbations, describing a pure-momentum transfer sourced by entropy. This term is exactly what appears in the effective sound speed  \cref{cs_de}, verifying that $\hat{c}_{s,\textrm{DE}}^2$ always differs from unity if there is a non-vanishing coupling current.
It is interesting to again note that the 
interacting models with derivative couplings must satisfy $h'(\mathcal{S}) = h'(0) = h_0$. A simple example that satisfies this criteria is linear functions $h(\mathcal{S}) \propto \mathcal{S}$, but the explicit form of $h$ does not affect the dynamics: only the value of the constant $h_0$.
Consequently, in the pure-derivative case, the only freedom is the value of the constant $h_0$ and the form of initial intrinsic entropy perturbation $\delta s(\vec{x})$, with the time-dependence coming solely from the potential $\hat{V}_{,\phi}$.

The dark matter density contrast evolution is governed by the perturbed continuity equation $ \nabla_{\mu} \delta T^{\mu}{}_{0} = - \delta J_{0}=0$, which is unchanged \cite{Ma:1995ey}
\begin{equation} 
    \delta'_{c} + (1+\frac{p_{c}}{\rho_{c}}) \big( \partial_{i} v^{i}_{c} - 3 {\Psi}' \big) + 3\mathcal{H} \Big(\frac{\delta p_{c}}{\rho_{c}} - \frac{p_{c}}{\rho_{c}} \delta_{c} \Big) = 0  \, . \label{cont1}
\end{equation}
The pressure perturbations $\delta p_c$ generally include non-adiabatic contributions \cite{Kodama:1984ziu}, which act as source terms for the evolution of the density perturbations (even in uncoupled models). However, we will shortly work with the entropic-CDM fluid, where the non-adiabatic pressure perturbations vanish like in standard cold dark matter models.

For the perturbed Euler equation there are new terms coming from $\nabla_{\mu}\delta T^{\mu}{}_{i} = - \delta J_{i}$, which lead to
\begin{equation}
    v^{i}_c{}'   + v^{i}_c \mathcal{H} - 3 \mathcal{H} \frac{{p_c'}}{{\rho'}} v^{i}_c  +\frac{\partial^{i}   \delta p_c}{p_c+\rho_c} + \partial^{i}   \Phi =\frac{\left( g_{,s} -h_0\hat{V}_{,\phi} \right) \partial^{i}  \delta s  }{\rho_c+p_c} \, ,
\end{equation}
where the shear stress $\sigma_c$ vanishes for our perfect fluid model.
Taking the spatial derivative $\partial_{i}$ and expanding in Fourier space with $\nabla^2  = -k^2 $ leads to
\begin{equation} \label{Eul1}
    {\theta}'_{c} + \theta_{c} \mathcal{H} - 3 \mathcal{H} \frac{p'_{c}}{\rho'_{c}} \theta_c  - k^2 \frac{ \delta  p_{c}}{p_{c}+\rho_{c}} - k^2 \Phi = - k^2 \frac{\left( g_{,s} -h_0\hat{V}_{,\phi} \right) \delta s  }{\rho_{c} + p_{c}} \, .
\end{equation}
In this equation, the modified entropy perturbation term looks exactly like a contribution to the DM non-adiabatic pressure perturbation or anisotropic stress \cite{Ma:1995ey,Ballesteros:2010ks,Kunz:2006wc,Malik:2004tf}, and similar to terms found in the GDM models \cite{Hu:1998kj,Kopp:2016mhm}.
We stress that the pressure perturbations appearing in the continuity equation, \cref{cont1}, do not have these new terms, and so the modification cannot simply be interpreted as non-adiabaticity. Similarly, the couplings do not contribute to anisotropic stress. Furthermore, the dark matter intrinsic entropy perturbations $\delta s(k)$ are explicitly coupled to dark energy, modulating a pure-momentum exchange between the two components.

The dark energy sector satisfies the same evolution equations as above, but with the source term in \cref{Eul1} taking the opposite sign, ensuring the overall conservation of energy-momentum. Consequently, the interaction primarily drives the dark energy velocity rather than its density, while the effective pressure response can differ markedly from that of a canonical scalar due to the modified effective sound speeds, see \cref{cs_de}. All other matter components are uncoupled and satisfy their usual conservation laws.

\subsubsection{Dynamical equations}

For the linear perturbations of the scalar field's equation of motion, we find additional source terms resulting from the entropy couplings 
\begin{equation}
    {\delta \phi}'' + 2 \mathcal{H} {\delta \phi}' +( k^2 +a^2 \hat{V}_{,\phi \phi}) \delta \phi = -2 a^2 \Phi \hat{V}_{,\phi} + {\phi}' ({\Phi}'+3{\Psi}') 
    - \left( a^2 g_{,s\phi}
    + k^2 h_0 \right) \delta s \, ,
    \label{eq:pert_kg}
\end{equation}
where we have expanded in terms of Fourier modes and used the background solutions in \cref{phi_back}. The new modified terms pick up specific scale dependences for the algebraic and derivative couplings, which we investigate in the next section.

Finally, we are equipped to explicitly compute the perturbed Einstein field equations, 
\begin{align} \label{EFE1}
    k^2 \Psi + 3 \mathcal{H} {\Psi}' +3 \mathcal{H}^2 \Phi &= -4\pi G \left( a^2 \sum_{c,b,r} \delta \rho_{I} + a^2 \delta \phi \hat{V}_{,\phi} - \Phi {\phi}'^2 + {\phi}' {\delta \phi}' + a^2 g_{,s}   \delta s \right)\, , \\ \label{EFE2}
    {\Psi}'' +  \mathcal{H} (2{\Psi}'+{\Phi}') + \mathcal{H}^2 \Phi + 2{\mathcal{H}}'\Phi + k^2 (\Psi - \Phi)  &= 4\pi G   \left(a^2 \sum_{c,b,r} \delta p_{I} - a^2 \delta \phi \hat{V}_{,\phi} - \Phi {\phi}'^2 +  {\phi}' {\delta \phi}' -  a^2 g_{,s} \delta s  \right)\, , \\
   k^2 ({\Psi}' + \mathcal{H} \Phi) &= 4\pi G   \left( a^2 \sum_{c,b,r} \theta_{I} (\rho_{I} + p_{I}) + k^2 {\phi}'  \big(h_0 \delta s + \delta \phi \big) \right) \, ,  \label{EFE3}
   \\
   k^2 (\Psi - \Phi) &= 12\pi G  a^2 \sum_{c,b,r}(\rho_I + p_I) \sigma_I \, , \label{EFE4}
\end{align}
where $(\rho_I + p_I) \sigma_I = -(\hat{k}_i \hat{k}_j - \frac{1}{3} \delta_{ij}) \Sigma^{i}_{I \,j}$ represents the anisotropic stress of the matter components\footnote{In our numerical implementation, $\sigma_I$ is obtained  from the standard \texttt{CLASS} Boltzmann hierarchies \cite{lesgourgues2011cosmic,Blas_2011,lesgourgues2011cosmic2}, and the entropy couplings do not introduce any new linear anisotropic stress, so the shear sector is unchanged.}, typically dominated by neutrinos \cite{Ma:1995ey}. For our entropic-CDM fluid, and for our effective dark energy sector, \cref{TDE}, the anisotropic stress vanishes identically.
The first two field equations are modified explicitly by the algebraic couplings while the time-space component is modified by the derivative interactions. Note that by using the definitions in \cref{rho_DE,p_DE,delta_DE,v_DE}, the above equations can be written with the sum $I$ including the dark energy terms too.

To directly see the effects of the coupling on the evolution of the metric perturbations, let us take the limit with vanishing anisotropic stress $\Phi=\Psi$ and a single effective matter-sector fluid component expanded into its adiabatic and non-adiabatic components, $\delta p = c_a^2 \delta \rho + \Gamma$ \cite{Kodama:1984ziu}. Combining equations \cref{EFE1,EFE2} then gives
\begin{multline} \label{EFE3d}
    {\Psi}'' + 3 \mathcal{H} (1+c_{a}^2) \Psi' + \left( 2 \mathcal{H}' +
     \mathcal{H}^2(1+3 c_{a}^2) +4 \pi G \phi'^2 (1-c_{a}^2) \right) \Psi + k^2 c_{a}^2 \Psi\\ = -4 \pi G  a^2 \left( (1+c_{a}^2) \delta \phi \hat{V}_{,\phi} + \frac{(c_{a}^2-1)}{a^2} \phi' \delta \phi'  + (1+c_{a}^2) g_{,s} \delta s - \Gamma
    \right) \, ,
\end{multline}
where $c_{a}$ and $\Gamma$ are the adiabatic sound speed and the non-adiabatic pressure perturbation of the effective matter fluid respectively, see \cref{Gamma}. In \cref{EFE3d} the algebraic couplings directly source the evolution of $\Psi$,
in addition to the usual scalar field terms and non-adiabatic pressure perturbations \cite{Celoria:2017xos,Malik:2004tf}. In particular, the term proportional to $g_{,s}\delta s$ enters the metric perturbation equation in the same way as the non-adiabatic pressure source $\Gamma$, showing that entropy perturbations in the dark sector act as an additional non-adiabatic source for the gravitational potential. 
Moreover, the derivative couplings can be seen explicitly when taking the dark energy rest frame $\delta \phi = - h_0 \delta s$, which again enter the RHS of \cref{EFE3d}.
Hence, both algebraic and derivative entropy couplings act as sources for the metric perturbations, directly modifying the gravitational dynamics beyond their indirect influence via the coupled dark matter sector.

\section{Application to entropic-CDM} \label{sec:entropic}
\subsection{Entropic-CDM fluid}
In order for these entropy interactions to couple to the scalar field and matter sectors, the dark matter energy density $\rho$ must depend on the entropy $s$. This precludes modelling dark matter in the traditional form $\rho=n m$ (see \cref{table} in \cref{sec:entropy}). Another requirement is that the matter fluid cannot simply be isentropic $s = \text{const.}$, which would lead to a vanishing entropy perturbation $\delta s =0$. We choose to study the simplest dark matter fluid model which can hold entropic interactions by working with a barotropic\footnote{See \cite{Ballesteros:2016kdx} for an overview using effective field theory of the allowed barotropic equations of state. Our solution, \cref{CDM_EoS}, can also be found in Ref.~\cite{Celoria:2017xos} for barotropic fluids that depend on two thermodynamic variables. Note that such solutions have not been studied in the context of dark matter and non-minimal couplings.
} equation of state $p = p(\rho)$. It automatically follows that the non-adiabatic pressure perturbations will vanish too \cite{MUKHANOV1992203}, see \cref{sec:entropy} for a detailed discussion.

We now choose to work with the following specific realisation of a non-isentropic but barotropic cosmological fluid which depends non-trivially on the entropy,
\begin{equation} \label{CDM_EoS}
    \rho_c(n,s) = m n  \gamma(s) \, ,
\end{equation}
where $m$ and $n$ are DM mass and particle number densities, and $\gamma(s)$ is an arbitrary function of the DM intrinsic entropy. We refer to this DM fluid as \textit{entropic-CDM} because the sound speed, isotropic pressure and pressure perturbations vanish 
\begin{equation}
    p_c = 0 \, , \quad c_{s,c}^2 = 0 \, , \quad \delta p_c = 0 \, . 
\end{equation}
However, the intrinsic entropy perturbations $\delta s$ still  affect the equations of motion explicitly through the non-minimal coupling, \cref{S_int}. Consequently, when couplings are turned off, the entropy perturbations completely decouple from the cosmological dynamics and the fluid behaves identically to standard CDM.  

For the entropic-CDM fluid, the intrinsic entropy perturbations $\delta s(k)$ must be set by primordial initial conditions and are frozen-in at all times. It should not, however, be confused with the standard relative isocurvature mode. Conventional isocurvature perturbations describe relative density fluctuations between different cosmological components and are associated with independent dynamical degrees of freedom with vanishing initial (adiabatic) curvature perturbation \cite{Kodama:1984ziu,Malik:2004tf}. In contrast, $\delta s$ represents an intrinsic entropy fluctuation within a single fluid component, which does not evolve dynamically and can be consistent with standard adiabatic initial conditions (vanishing isocurvature). Nonetheless, it is useful to note the analogous roles played by intrinsic entropy and relative isocurvature perturbations: both can source non-adiabatic pressure perturbations\footnote{For our specific entropic-CDM \cref{CDM_EoS}, the non-adiabatic pressure perturbations vanish identically, see \cref{sec:entropy}.}, and both correspond to independent primordial degrees of freedom whose amplitudes are specified by initial conditions \cite{Malik:2008im}. In \cref{sec:iso} we give a formal discussion of isocurvature and show that it can be set to zero in the presence of our entropic-CDM fluid and couplings, which we utilise in the next section.

Although we remain agnostic about the microscopic origin of the intrinsic entropy perturbation, it is useful to comment on possible early-Universe mechanisms which could give rise to the initial condition $\delta s(k)$. A possible origin is provided by light spectator fields, or more generally, entropy degrees of freedom during multi-field inflation, whose fluctuations can give rise to primordial entropy perturbations \cite{Gordon:2000hv}. If these fields contribute to the dark matter sector after inflation, then the effective CDM fluid can pick up intrinsic entropy perturbations. For instance, if $\chi$ denotes the relevant inflationary degree of freedom then one may write $s(\mathbf{x}) = s[\chi(\mathbf{x})]$ and $\delta s \propto s_{,\chi}\delta \chi$. In the entropic-CDM model described by \cref{CDM_EoS}, the dependence on $\chi$ would enter the effective fluid through the entropy-dependent function $\gamma(s)$. The precise inflationary model, its interactions with the dark sector and the underlying dark matter production mechanism will then determine whether the entropy perturbations $\delta s$ survive. For instance, these perturbations can survive provided that after CDM production any number-changing interactions, thermalisation with the visible sector, or other entropy-producing processes in the dark sector remain inefficient enough to preserve the spatial dependence of $s$. 
This is analogous to the usual statement that CDM isocurvature depends on the production history of the CDM component \cite{Lyth:2003ip}, while here the remnant is interpreted as an intrinsic entropy perturbation of the effective dark matter fluid \cite{Celoria:2017bbh}.

\subsection{Continuity and Euler equations}
Working with the entropic-CDM fluid, \cref{CDM_EoS},
the continuity equation, \cref{cont1}, becomes
\begin{equation}  \label{DM_delta}
    {\delta}'_{c} + \theta_{c} - 3 {\Psi}' = 0 \, ,
\end{equation}
and the Euler equation, \cref{Eul1}, is
\begin{equation} \label{DM_theta}
    {\theta}'_{c}   + \theta_{c}  \mathcal{H} -  k^2 \Phi = -k^2Q_s \delta s(k)  \, , \quad \text{with} \quad  Q_s(a) := \frac{g_{,s} - h_0 \hat{V}_{,\phi}}{\rho_c} \, ,
\end{equation}
where $Q_s(a)$ is the coefficient of a new time-dependent source term resulting from the non-minimal coupling. This term comes directly from the coupling current, \cref{Jpert}, and modulates the strength of the momentum-exchange interactions.
The source term vanishes when the coupling is switched off, or when the fluid is isentropic $\delta s = 0.$
We have also written the intrinsic entropy perturbation explicitly in Fourier space, with $\delta s(k)$ labelling the time-independent spatial mode. Unlike the Type-3 or elastic scattering models \cite{Linton:2021cgd}, our modification does not contribute as a friction term for $\theta_c$ but instead introduces an inhomogeneous source or forcing term. Moreover, this has a natural scale dependence which is not easily obtained in other models.

Lastly, let us differentiate \cref{DM_delta} and use \cref{DM_theta} to obtain a second-order equation for the growth of dark matter perturbations 
\begin{equation}
    {\delta}''_{c} +  \mathcal{H} {\delta}'_{c} -3 {\Psi}' \mathcal{H} - 3 \Psi'' + k^2 \Phi = k^2 Q_s \delta s(k)  \, ,
    \label{deltacpp}
\end{equation}
which can be compared to the uncoupled case with non-adiabatic pressure perturbations \cite{Dent:2008ek}. 
In the next section, we will study this equation analytically by taking the matter-dominated quasi-static limit and using the Poisson equation to eliminate the $\Psi$ terms. However, even at this stage we notice that the algebraic and derivative couplings both act as a source of the evolution of dark matter perturbations, and this will be reflected in the rate of structure growth.

Unlike a conventional scalar fifth force, the new term in \cref{deltacpp} is not proportional to $\delta_c$ and hence cannot be absorbed into a rescaling of Newton's constant. This will be shown explicitly in the quasi-static limit below. Instead, it acts as an inhomogeneous source fixed by the primordial entropy perturbation. Its sign determines whether the sourced solution adds constructively or destructively to the standard growing mode, so a suppression of the total matter perturbation is not, by itself, a diagnostic of a repulsive force or of a gradient instability. In the next section we numerically verify that the dark sector perturbations remain stable over the regimes of interest.

\subsection{Quasi-static limit} \label{sec:quasi}

The quasi-static limit corresponds to the sub-horizon, slowly-varying regime of linear perturbations in which spatial gradients dominate over time derivatives. In the matter-dominated epoch, the effects of anisotropic stress can be neglected from \cref{EFE4}, leading to $\Phi\simeq\Psi$. For modes deep inside the horizon, $k \gg a H = \mathcal{H}$, the metric potentials and scalar field perturbations vary on Hubble timescales, while gradient terms scale as $k^2$.
This means that under this approximation we can neglect derivative terms $\Psi',\Psi'',\delta \phi',\delta \phi''$ (and any other terms suppressed by $(aH/k)^2$), while retaining the background time dependence. 
In this regime, the Poisson equation, \cref{EFE1}, becomes algebraic in $\Psi$ while the Klein-Gordon equation, \cref{eq:pert_kg}, becomes algebraic in $\delta \phi$.
This approximation is expected to hold for modes sufficiently inside the horizon and at late times, i.e., not near horizon-crossing or when rapid background evolution makes time derivatives large compared to $k^2$ terms. 

From the perturbed Klein-Gordon equation, \cref{eq:pert_kg}, (dropping $\delta \phi'', \delta \phi', \Psi'$) we obtain
\begin{equation}
    (k^2 + a^2 m_{\rm eff}^2) \delta \phi \simeq -2 a^2 \Psi \hat{V}_{,\phi} - (a^2 g_{,s\phi} + k^2 h_0) \delta s \quad \Longrightarrow \quad \delta \phi \simeq - \frac{2 a^2  \hat{V}_{,\phi} \Psi + (a^2 g_{,s\phi} + k^2 h_0) \delta s}{k^2 + a^2 m_{\rm eff}^2} \, ,
    \label{eq:deltaphiqs}
\end{equation}
with $m_{\rm eff}^2 \equiv \hat{V}_{,\phi \phi}$.
In the Poisson equation, \cref{EFE1}, neglecting time-derivatives and terms suppressed by $(aH/k)^2$ leads to 
\begin{equation} \label{Poisson}
    k^2 \Psi \simeq - 4 \pi G a^2 \left( \delta \rho_m +  \Delta \rho_s \delta s \right) \quad \text{with} \quad \Delta \rho_s(k,a) := g_{,s}  - \frac{a^2 g_{,s\phi} + k^2 h_0 }{k^2 + a^2 m_{\rm eff}^2}  \hat{V}_{,\phi} \, ,
\end{equation}
where at late times we approximate $\delta \rho_m \approx \delta \rho_c$ (baryons and radiation are subdominant in the matter density).
We have also defined a new scale- and time-dependent effective source term $4\pi G a^2 \Delta \rho_{s} \delta s$, which depends on the background scalar field dynamics and the type of entropy coupling. 
In the quasi-static limit, the equation for the growth rate, \cref{deltacpp}, becomes simply
\begin{equation}
    \delta_c'' + \mathcal{H} \delta_c' + k^2 \Psi \simeq k^2 Q_s \delta s  \, ,
\end{equation}
where $k^2 Q_s \delta s$ is the modified source term of the Euler equation, defined in \cref{DM_theta}.
Finally, eliminating $\Psi$ using \cref{Poisson}, yields the growth-rate equation in the quasi-static regime
\begin{equation}
    \delta_c'' + \mathcal{H} \delta_c' - 4 \pi G a^2 \left[ \rho_c \delta_c + \Delta \rho_s(k,a) \delta s \right]  \simeq  k^2 Q_s (a) \delta s  \, .
    \label{eq:deltacpp}
\end{equation}

The intrinsic entropy perturbation contributes through two distinct channels: i) as an effective density source $4\pi G a^2 \Delta \rho_s(k,a) \delta s$ in the Poisson equation and ii) as a direct ``fifth-force"-like source $k^2 Q_s (a) \delta s$ entering the momentum equation. The relative importance of these two entropy-induced terms depends on cosmic time, scale, and the type of coupling (algebraic or derivative). In general, these source terms cannot be absorbed into a simple rescaling of Newton's gravitational constant $G$ for dark matter. This differs from models in which the interaction can be recast as a scale-independent $G_{\rm eff} (a)$ \cite{Tsujikawa:2007gd,Amendola:2003wa} or as a modified friction term $\mathcal{H}_{\rm eff} (a)$ multiplying $\delta_c'$ \cite{Baldi:2014ica,Simpson:2010vh,Poulin:2022sgp}. Instead, $\delta s$ is an independent degree of freedom, so the growth equation remains inhomogeneous and entropy perturbations act as an external source for $\delta_c$. Moreover, the new terms are generically scale dependent, with both $\Delta \rho_s(k,a)$ and $\delta s(k)$ depending on mode $k$, and a factor of $k^2$ naturally appearing in the Euler source term $ \propto k^2 Q_s$. We therefore generically expect scale-dependent signatures, depending on the explicit parametrisation of $\delta s(k)$.

Before going into more details about the impact of this entropy coupling, it is instructive to illustrate the relative effects of these contributions. To that end, we study two simple toy-models\footnote{Motivation for choosing these two models is given shortly, below \cref{eq:models}.}, $g = g_0 \phi s$ and $h = h_0 \nabla_{\mu} \phi \nabla^{\mu} s$, where $g_0$ and $h_0$ are constants that define the strength of the coupling. 
For the pure-derivative coupling ($g_{0}=0$), the coefficients of the sources reduce to
\begin{equation}
    Q_s^{\rm der} (a) = - \frac{h_0 }{\rho_c} \hat{V}_{,\phi} \quad \text{and} \quad \Delta \rho_s^{\rm der} (k,a) = - \frac{ k^2 h_0 }{k^2 + a^2 m_{\rm eff}^2}  \hat{V}_{,\phi}\, ,
    \label{eq:qs_der}
\end{equation}
while for the pure-algebraic case ($h_0=0$) we have
\begin{equation}
   Q_s^{\rm alg} (a) = \frac{g_{0} \phi}{\rho_c}  \quad \text{and} \quad \Delta \rho_s^{\rm alg} (k,a) = g_{0} \phi  - \frac{a^2 g_{0} }{k^2 + a^2 m_{\rm eff}^2}  \hat{V}_{,\phi}\, .
   \label{eq:qs_alg}
\end{equation}
We take a standard effective exponential potential $\hat{V} = V_0 \exp(-\lambda \phi/{\rm M}_{\rm Pl})$, with parameters and initial conditions discussed in the next section. 
Since we expect prominent effects of the modifications to occur on small scales, the quasi-static expressions in \cref{eq:qs_der,eq:qs_alg} provide a representative approximation of the impact of the entropy coupling on these scales.

\begin{figure}
      \subfloat{\includegraphics[height=0.35\linewidth]{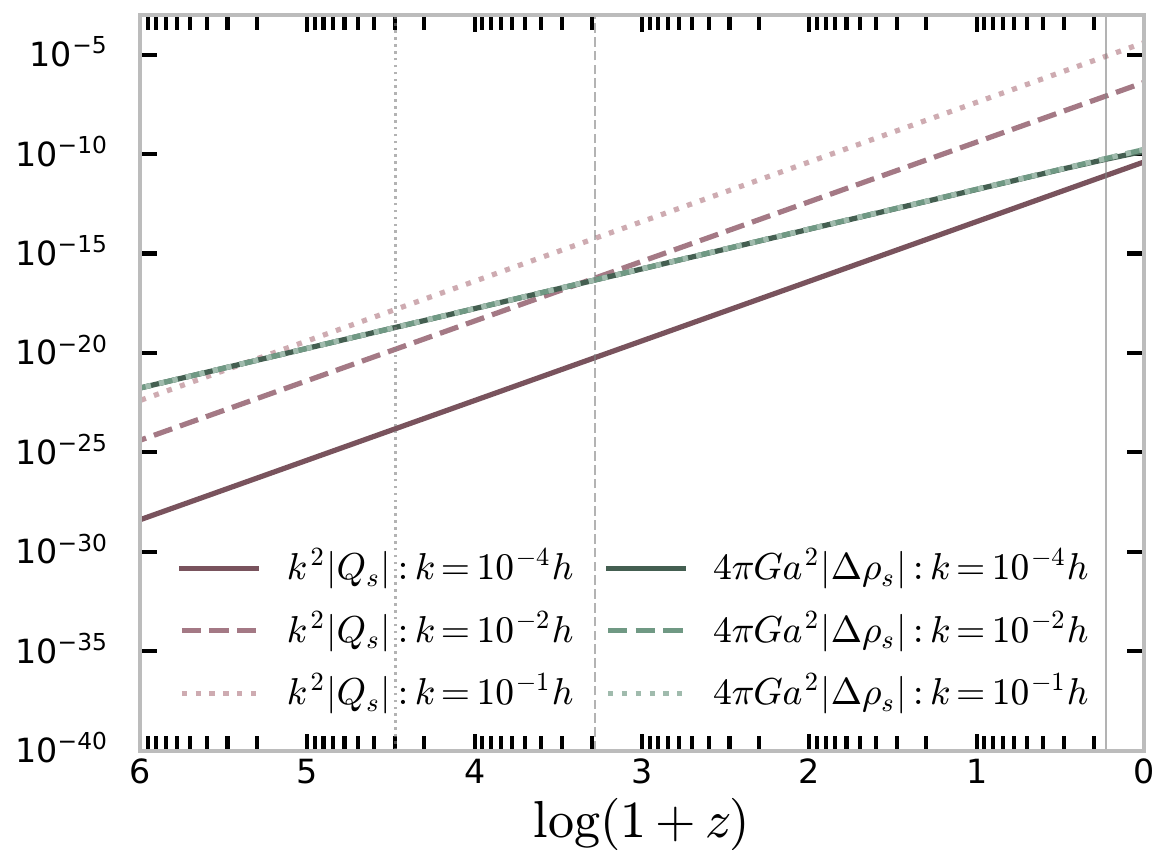}}
      \qquad
      \subfloat{\includegraphics[height=0.35\linewidth]{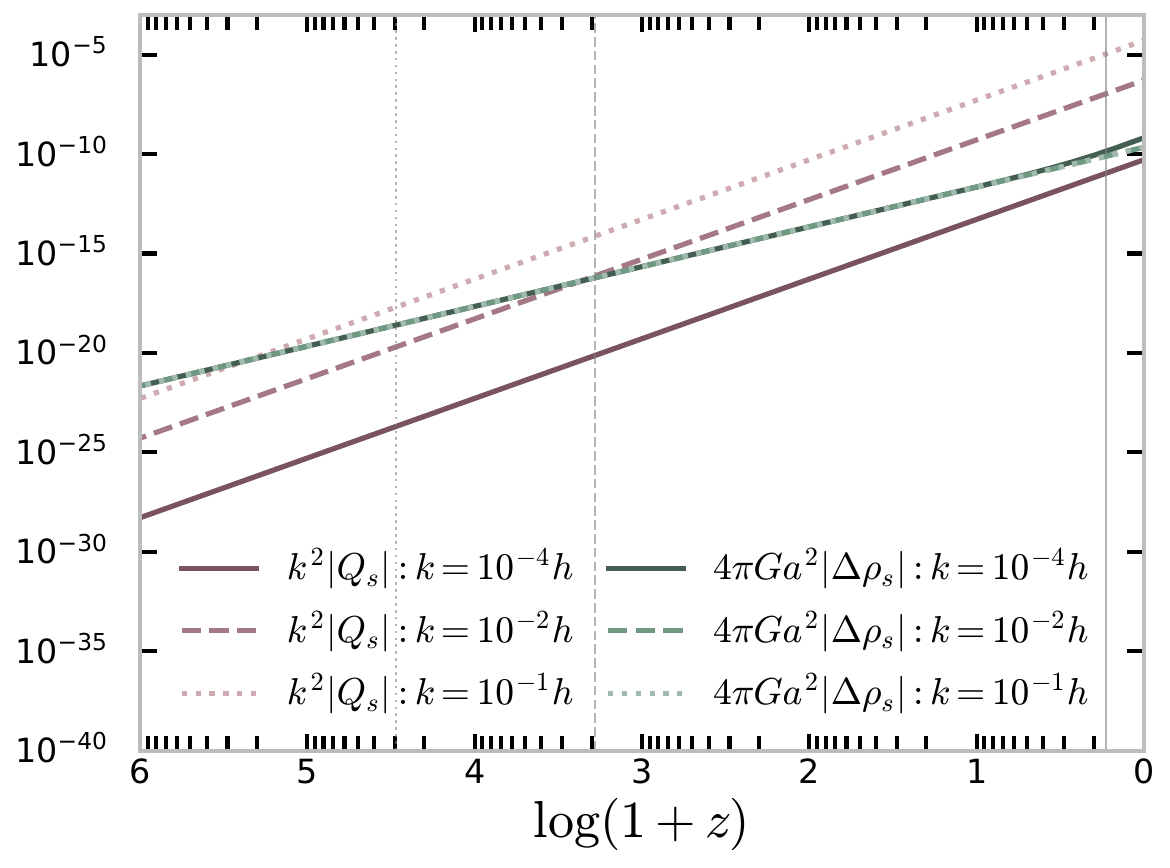}}
  \caption{Comparison of the entropy-induced source terms entering the dark matter growth equation in the quasi-static limit, \cref{eq:deltacpp}, for the derivative coupling (left) and algebraic coupling (right), according to the definition in \cref{eq:models}. We show the Euler-source contribution $k^2|Q_s|$ (rose) and the effective density contribution $4\pi G a^2 |\Delta \rho_s|$ (green) for three representative modes, $k=10^{-4}h\,{\rm Mpc}^{-1}$ (solid), $k=10^{-2}h\,{\rm Mpc}^{-1}$ (dashed), and $k=10^{-1}h\,{\rm Mpc}^{-1}$ (dotted). The vertical lines mark horizon crossing, $k=aH$, for each mode. The effective density source is scale independent for most of the evolution, except for the largest scales ($k=10^{-4}h\,{\rm Mpc}^{-1}$) upon horizon entry.
  }
  \label{fig:comps}
\end{figure}

In \cref{fig:comps} we compare $k^2 |Q_s|$ (rose) and $4 \pi G a^2 |\Delta \rho_s|$ (green) for representative scales $k$, with the vertical lines indicating horizon entry for each mode ($k=aH$).
In the derivative case, the effective density source $\Delta \rho_s^{\rm der}$ has a Yukawa-type structure $\propto k^2/(k^2+a^2 m_{\rm eff}^2)$, leading to a suppression in the large-scale regime $(k \ll a m_{\rm eff})$ or approximately scale-independent behaviour on sub-Compton scales $(k \gg a m_{\rm eff})$. In the algebraic case, $\Delta \rho_s^{\rm alg}$ contains a scale-independent term $g_0 \phi$ plus a Yukawa-type correction $\propto (k^2+a^2 m_{\rm eff}^2)^{-1}$, so that for $(k \ll a m_{\rm eff})$ the dominant contribution becomes approximately scale independent.
For the exponential potential considered here, we typically have $m_\textrm{eff} \sim \mathcal{O}(H)$, and the Compton scale is comparable to the horizon scale. The quasi-static approximation applies only to sub-horizon modes, most of which are also in the sub-Compton regime $k \gtrsim  a m_{\rm eff}$ over most of the expansion history. 
Indeed, only on the largest scales ($k=10^{-4}h$ in \cref{fig:comps}) do we observe a mild suppression or enhancement of $\Delta \rho_s^{\rm der}$ and $\Delta \rho_s^{\rm alg}$ for modes that have recently entered the horizon close to $z=0$. However, these near-horizon scales are outside of the regimes of validity for the quasi-static limit. Therefore, on most scales of interest, we see an approximately scale-independent effective density source $\Delta \rho_s$ in both the algebraic and derivative cases.

In contrast, the Euler-source contribution scales with $k^2$ in both the algebraic and derivative cases, and grows at late times due to the background time-dependence of $Q_s \propto  \hat{V}_{, \phi} \rho_c^{-1}$ or $Q_s \propto  \phi \rho_c^{-1}$.
We observe that upon horizon entry, the magnitude of the source contributions $k^2|Q_s|$ and $4 \pi G a^2 |\Delta \rho_s|$ are comparable to one another. However, once a mode is safely inside the horizon ($k \gg a H$),
the fifth-force–like contribution typically overtakes the effective density term  and provides the dominant entropy-driven modification on all but the largest scales of \cref{fig:comps}. 
We have verified that this behaviour holds irrespective of the size of the coupling constants $g_0$ and $h_0$, making this result general within the quasi-static regime. 
In \cref{sec:obs} we will consequently see that these entropy couplings give rise to considerable deviations from the standard $\Lambda$CDM growth history at low redshifts.

\subsection{Large-scale behaviour} \label{sec:large}
While the small-scale modifications arising from the fifth-force term in the Euler equation, \cref{DM_delta}, are readily understood within the quasi-static approximation, it is less obvious that the coupling can also affect perturbations on very large scales. This behaviour follows from the role of the entropic-CDM velocity divergence in the momentum constraint, \cref{EFE3}, which governs the time evolution of the metric potentials. More concretely, 
\begin{equation}\label{EFE0i}
     {\Psi}' + \mathcal{H} \Phi = \frac{4\pi G a^2}{k^2} \sum_{c,b,r,\textrm{DE}} \theta_{I} (\rho_{I} + p_{I}) \, ,
\end{equation}
where the summation now includes the dark energy terms $\rho_{\textrm{DE}}$, $p_{\textrm{DE}}$ and $\theta_{\textrm{DE}} $. Given that the background is unmodified, and that we expect the dark energy sector to account for accelerated expansion at late times, $w_{\textrm{DE}} \approx -1$, its contribution to the total momentum density $(1+w_{\rm DE})\rho_{\rm DE}\theta_{\rm DE}$ is suppressed due to $1+w_{\rm DE} \ll 1$. This implies that the modifications to the RHS of the momentum constraint, \cref{EFE0i}, predominantly come from the changes to the dark matter velocities\footnote{We have indeed verified numerically that, for the models studied in \cref{sec:obs}, the dominant contribution to both the Poisson equation, \cref{EFE1}, and momentum constraint, \cref{EFE3}, comes solely from the entropic dark matter terms, with dark energy perturbations being heavily suppressed.}. 
The velocity of the entropic-CDM fluid obeys its modified Euler equation, \cref{DM_theta}, with an additional scale-dependent source $k^2 Q_s \delta s$. 
However, the velocity divergence sources the metric potentials in the momentum constraint, \cref{EFE0i}, with a suppression factor of $1/k^2$. Consequently, the explicit scale dependence of the fifth-force in the Euler equation does not necessarily translate into a suppression of the impact of the coupling on the time evolution of the metric potentials on large scales. Once $Q_s$ becomes significant at late times, the coupling can therefore modify the large-scale evolution of the metric potentials.

This effect will be most prominent for modes close to horizon entry, where changes to the evolution of the metric potentials can affect structure growth through the continuity equation, \cref{DM_delta}, and give rise to a late-time integrated Sachs–Wolfe effect. On the other hand, for modes inside the horizon $(k \gg a H)$ the metric potentials evolve slowly during matter domination, so the $\Psi'$ terms in the continuity equation are negligible and growth is determined primarily by the Euler and Poisson equations. Therefore, we expect that entropy couplings will play a non-trivial role on both large scales (via the momentum constraint) and small scales (via the direct fifth-force term in the Euler equation), depending on the scale dependence of $\delta s(k)$. 

Interestingly, the relative effects within these two regimes naturally oppose one another: a suppression of growth on small scales can be accompanied by an enhancement on very large scales, and vice versa. On small scales, where $k^2 Q_s \delta s$ is large, the dominant effect arises from the modified Euler equation, where the coupling directly alters the dark matter velocities and can lead to a suppression (or enhancement) of structure growth. On near-horizon scales, however, the modified velocity divergence feeds into the momentum constraint and significantly alters the evolution of the metric potentials, which then affects density growth through the continuity equation. This can then lead to a corresponding enhancement (or suppression, respectively) of power. 
Similar behaviours have been identified in other interacting dark sector models with momentum transfer, including the Type-3 models of Refs.~\cite{Pourtsidou:2013nha,Linton:2021cgd}, though the effects are qualitatively different. Moreover, the mechanism responsible for this behaviour in these entropy-coupled models, with the Euler-source term $k^2 Q_s \delta s$ entering through two distinct channels, has not been discussed before in these contexts. This behaviour should be clearly imprinted on the matter spectrum, which we compute in the next section.

\section{Impact on observables} \label{sec:obs}

In this section we investigate the observational consequences of the interacting entropic dark sector models with algebraic and derivative couplings. We compare these models to the uncoupled quintessence case ($\phi$CDM) and to $\Lambda$CDM, focusing on modifications to the matter power spectrum, the CMB lensing potential, and the CMB spectrum of temperature anisotropies. 

To study the cosmological implications of each class of entropy couplings we use a modified version of the Einstein-Boltzmann solver \texttt{CLASS}\footnote{\href{https://github.com/elsateixeira/class_entropy}{https://github.com/elsateixeira/class\_entropy}}~\cite{lesgourgues2011cosmic,Blas_2011,lesgourgues2011cosmic2}, adapted to include entropic-CDM coupled to scalar field dark energy. This requires specifying the statistical properties of the intrinsic entropy perturbations. In a concrete microphysical model, these would be determined by the underlying dark matter physics, for instance, through additional light degrees of freedom during inflation \cite{Cicoli:2021itv} or non-equilibrium processes such as phase transitions \cite{Lyth:2009imm}. For simplicity, in the absence of a unique microscopic prescription, we adopt a phenomenological approach and parametrise the entropy perturbations in terms of their stochastic amplitude in Fourier space.
More precisely, we assume that $\delta s(\vec{x})$ is a statistically homogeneous and isotropic random field, with vanishing cross-correlation with the curvature perturbation $\langle \zeta_k \, \delta s_{\mathbf{k}'}^{*} \rangle =0$. Its Fourier modes are characterised by the power spectrum $P_s(k)$ defined through
\begin{equation}
\langle \delta s_{\mathbf{k}} \, \delta s_{\mathbf{k}'}^{*} \rangle
=
(2\pi)^3 \delta^{(3)}(\mathbf{k}-\mathbf{k}')\, P_s(k)\, ,
\end{equation}
where $k \equiv |\mathbf{k}|$.
Since the intrinsic entropy perturbations are non-dynamical and conserved along the fluid flow, their stochastic properties are fully specified once their primordial spectrum is chosen.
We adopt the following power spectrum\footnote{Note that in practice, $\delta s$ is normalised to the curvature fluctuations just like any other perturbed quantity in \texttt{CLASS}. The entropy power spectrum, \cref{eq:deltas_def}, is therefore normalised relative to the primordial power spectrum of curvature fluctuations $\mathcal{P}_{\mathcal R}(k)$.} for the intrinsic entropy degrees of freedom
\begin{equation}
\mathcal P_s(k) \equiv \frac{k^3}{2\pi^2} P_s(k) = \mathcal{A}_{\textrm{e}}\left(\frac{k}{k_{p}}\right)^n  \exp \left[-\left(\frac{k}{k_c}\right)^{p_c}\right] \, , 
\label{eq:deltas_def}
\end{equation}
where $\mathcal A_{\textrm{e}}$ sets the amplitude at the pivot scale $k_{p}$, $n$ controls the tilt, $k_c$ is a phenomenological UV cut-off scale and $p_c$ determines the sharpness of the cut-off. 
We adopt this form for the entropy spectrum to regularise the high-$k$ behaviour, preventing the unphysical growth that would arise from extrapolating a pure power-law to arbitrarily small scales ($k\gtrsim1\, \text{Mpc}^{-1}$). In particular, the dominant entropy contribution to the DM growth equation enters through the combination $k^2 Q_s(a)\,\delta s(k)$, so that for a non-negative tilt\footnote{We also noted in \cref{sec:large} that large-scale effects are induced by the coupling via the momentum constraint, \cref{EFE0i}. A negative tilt would overly enhance these effects.} ($n\geq 0$), a power spectrum scaling as $\mathcal{P}_{s} \propto k^n$ would over-weight increasingly small scales. For this reason, we introduce an exponential suppression $\exp\left[-(k/k_c)^{p_c}\right]$ which regulates the UV behaviour and should be viewed as encoding either the finite range of validity of the effective description, or the presence of microphysical damping beyond a characteristic scale. Moreover, given that the entropy is an effective fluid variable in the EFT viewpoint, the necessity of a cut-off is expected \cite{Ivanov:2022mrd}. In practice, this implies that entropy-induced deviations saturate for $k\gtrsim k_c$, rather than continuing to grow with $k$. A UV cut-off is therefore a standard phenomenological choice for models which introduce new stochastic primordial fields that affect small-scale structure, often seen in certain isocurvature models \cite{Feix:2019lpo,Chung:2023xcv} or models of primordial magnetic fields \cite{Kahniashvili:2005xe,Kunze:2011bp,Ralegankar:2024ekl,Cruz:2023rmo}. For the illustrative cases considered, we choose the characteristic cut-off scale $k_c = 1\, \text{Mpc}^{-1}$ which is successful in taming unphysical small-scale behaviour. 
 
In the particular case $n=0$, the entropy perturbations give an approximately white-noise-like spectrum (on top of the standard spectrum of adiabatic fluctuations) for $k\ll k_c$, characterised by equal power and vanishing spatial correlations on these scales.
Moreover, if these perturbations are seeded during inflation one expects a nearly scale-invariant spectrum \cite{Guth:1980zm}, analogous to that of light spectator fields.
This choice provides a particularly simple and well-motivated baseline, which we adopt for most of our analysis. 
In summary, hereafter we assume $\mathcal{A}_{\rm e}=1$, $n=0$, $k_p=0.05\,\mathrm{Mpc}^{-1}$, $k_c=1\,\mathrm{Mpc}^{-1}$, and $p_c=2$, unless otherwise stated.

For the purpose of the numerical analysis we consider the following representative choices for the algebraic and derivative coupling functions:
\begin{equation} \label{eq:models}
 g(\phi,s) = g_0 \phi s\quad \text{and}\quad    h(\nabla_{\mu} \phi \nabla^{\mu} s) = h_0 \nabla_{\mu} \phi \nabla^{\mu} s \, ,
\end{equation}
where $g_0$ and $h_0$ are constants with units of mass$^3$ and mass, respectively. As discussed below \cref{Jpert}, the derivative models are essentially fixed to be of this form since $\nabla_{\mu}\phi \nabla^{\mu}s$ vanishes at linear order. For the algebraic models, we choose the simplest case of a coupling linear in both $\phi$ and $s$. Therefore the strength of the coupling will be determined by the constants $h_0$ and $g_0$ (degenerate with $\mathcal{A}_{\rm e}$, which has been set to 1 without loss of generality).
We assume that the dark energy scalar field is evolving in an effective exponential self-interacting potential of the form 
\begin{equation} \label{eq:V}
 \hat{V}(\phi) = V_0 e^{-\lambda \phi/{\rm {\rm M}_{\rm Pl}}} \, ,   
\end{equation}
where $V_0$ is a constant scale with units of mass$^4$ and $\lambda$ is a dimensionless parameter defining the slope of the potential. This choice allows for scaling solutions and has been thoroughly studied in the context of quintessence models \cite{Copeland:1997et}. For the numerical study presented below we take a nearly flat potential with $\lambda=0.1$ and $\phi_i=1\, {\rm M}_{\rm Pl}$ and a small negative initial kick $\phi_i'=-8 \times 10^{-12}\, {\rm M}_{\rm Pl}/\, {\rm Mpc}$, to isolate the effects arising from the entropy coupling\footnote{We have verified that the inclusion of the power-law contribution to $\hat{V}$ coming from the algebraic coupling $g_0 \phi$ does not lead to any significant changes to the overall evolution of the cosmological background or perturbations. 
We therefore choose to absorb the background contribution $f(\phi,s,0)$ into the definition of the effective scalar-field potential, leading to \cref{eq:V}.}. 

Finally, throughout this study we impose adiabatic initial conditions for the cosmological perturbations. Although dark-sector interactions can in general affect the viability of adiabatic initial conditions \cite{Majerotto:2009np}, in our framework the adiabatic growing mode can still be constructed consistently once the modified entropy-sourced terms are taken into account. This is because the couplings leave the background evolution unchanged and do not significantly alter the early-time behaviour of the perturbations.
The corresponding initial conditions are derived in \cref{sec:initial}, where the standard adiabatic solution is supplemented by the contributions induced by the intrinsic entropy perturbations and then implemented in \texttt{CLASS}. We verify numerically that these initial conditions lead to stable and well-behaved solutions across the parameter space explored, and that in the limit of vanishing coupling the early-time evolution reduces to the standard adiabatic case (see \cref{fig:condinis} in \cref{sec:initial}).

\subsection{Impact on dark matter perturbations}

\begin{figure}
      \subfloat{\includegraphics[width=0.48\linewidth]{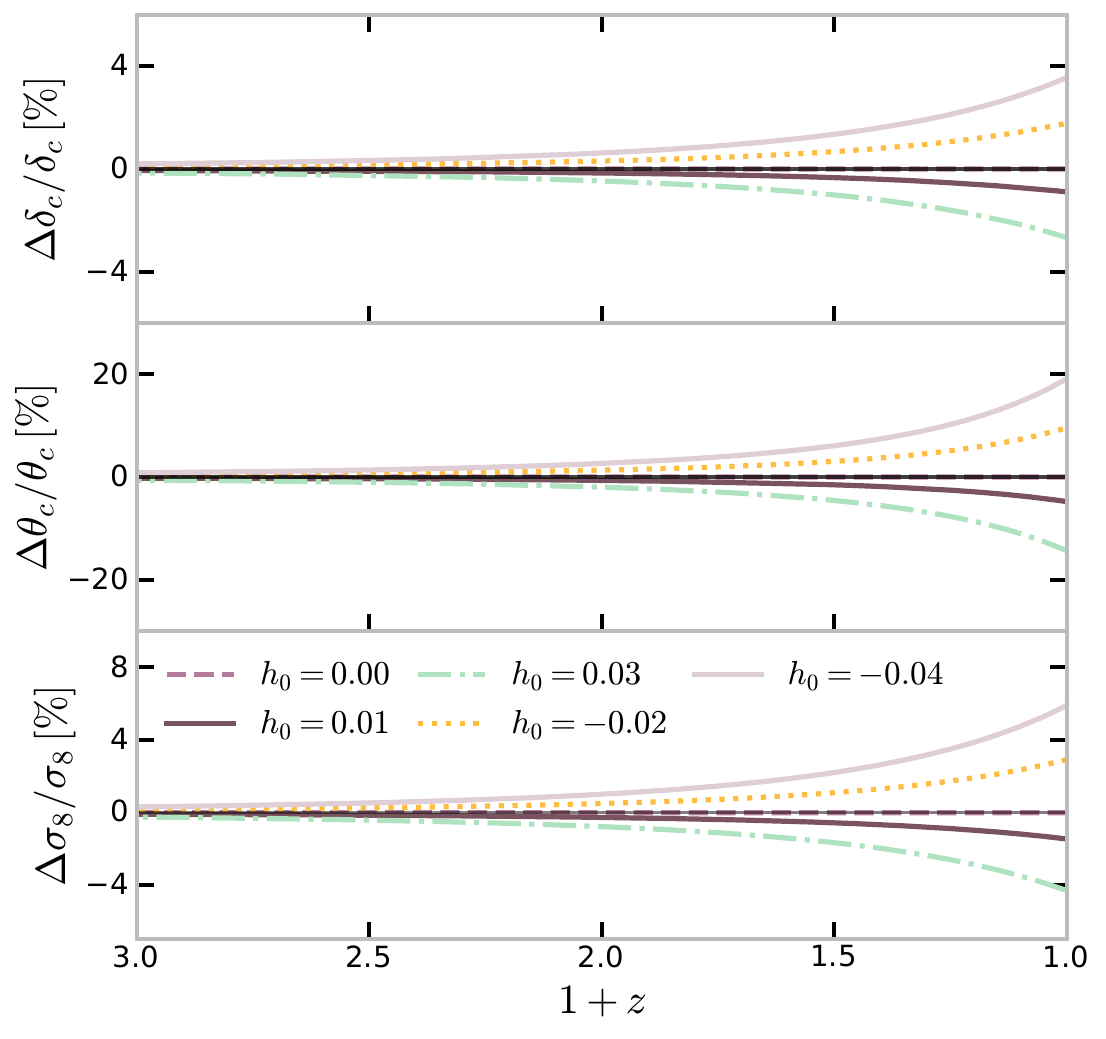}}
      \hfill
      \subfloat{\includegraphics[width=0.48\linewidth]{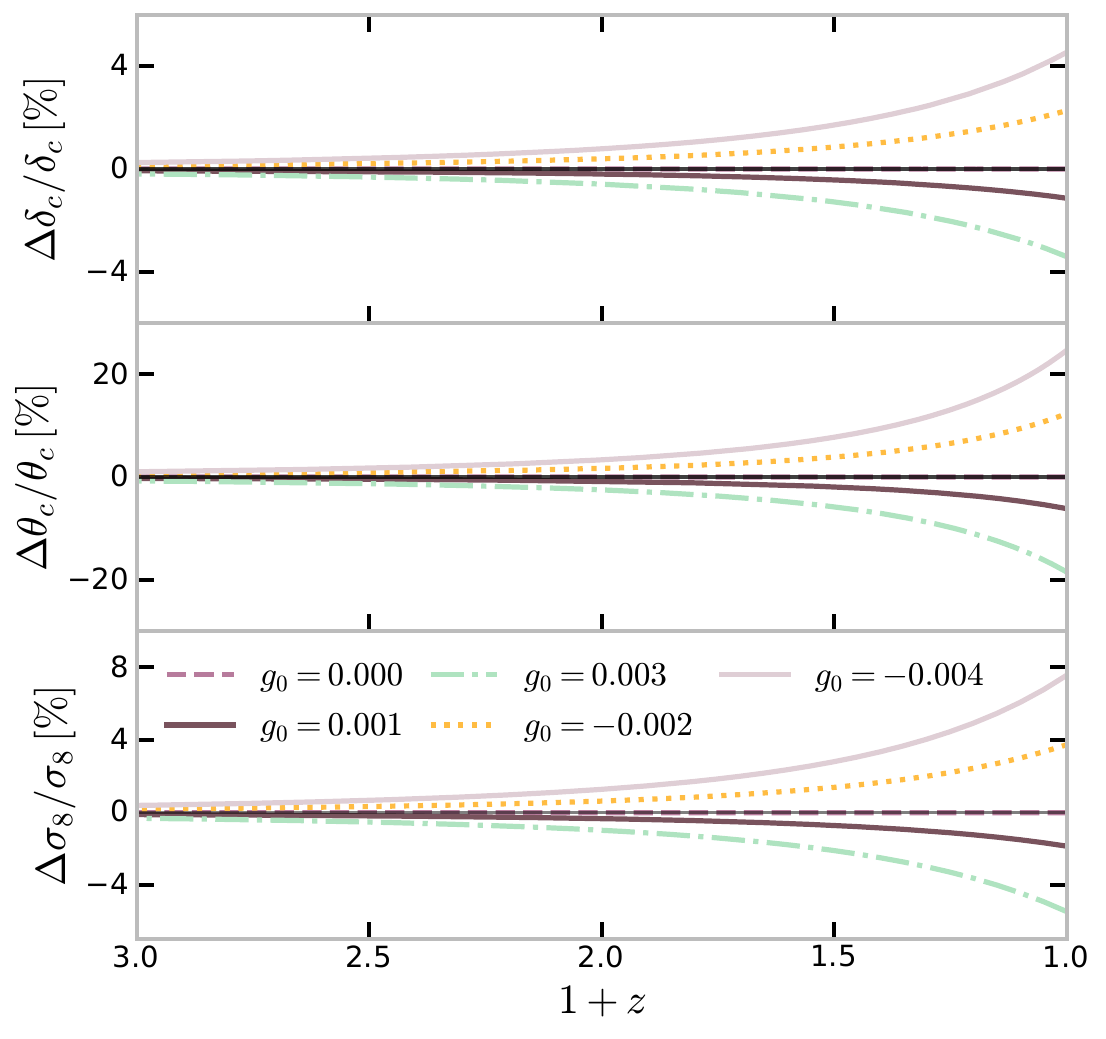}}
  \caption{Relative deviations of the entropic-CDM perturbations and clustering amplitude with respect to $\Lambda$CDM and as a function of redshift $z$ for $k_8=0.125\,h\,{\rm Mpc}^{-1}$. The top and middle panels show the fractional changes in the dark matter density contrast $\delta_c$ and velocity divergence $\theta_c$, while the bottom panels show the relative change in $\sigma_8$. The entropy perturbation $\delta s$ is parameterised as in \cref{eq:deltas_def}. The left panels show the derivative coupling for different values of $h_0$, while the right panels depict the algebraic coupling with varying $g_0$, as defined in \cref{eq:models}. In both cases the effects remain negligible at early times and grow only at late times, when the entropy coupling becomes dynamically relevant. The largest deviations occur in $\theta_c$, reflecting the fact that the interaction enters directly through the Euler equation, \cref{Eul1}, while the impact on $\delta_c$ and $\sigma_8$ is induced indirectly through the changes to the $\theta_c$ and the metric potentials.}
  \label{fig:delta_theta_DM}
\end{figure}

Before turning to the cosmological observables, it is useful to consider the direct impact of the coupling on the dark matter perturbations. In \cref{fig:delta_theta_DM} we show the relative deviations in $\delta_c$ (top panels), $\theta_c$ (middle panels), and $\sigma_8$ (lower panels) with respect to $\Lambda$CDM, for the intermediate scale $k_8=0.125\,h\,{\rm Mpc}^{-1}$. In both the derivative and algebraic models the coupling becomes relevant only at late times, and consequently the deviations in the perturbations remain negligible at early times. As expected, the effect is much more pronounced in the velocity divergence than in the density contrast, reflecting the fact that the interaction enters directly in the Euler equation through the fifth-force term $k^2 Q_s(a) \delta s$, and only indirectly modifies $\delta_c$ through the subsequent evolution. The same late-time behaviour is inherited by the  matter fluctuation amplitude $\sigma_8$, which gives an integrated measure of the modifications to the growth history. We see that positive values of the coupling parameters suppress structure growth, while the opposite holds for negative values. This suggests that such couplings may play a role in explaining the (weak) tension between CMB and weak-lensing determinations of the $S_8\equiv \sigma_8\sqrt{\Omega_m/0.3}$ parameter \cite{Heymans:2020gsg,Wright:2025xka,DES:2025tna}. Lastly, the two classes of models (algebraic and derivative) display the same qualitative behaviour, differing mainly in the detailed scale dependence of the source terms which, as discussed in \cref{sec:quasi}, is expected to be mild for the ranges of scales considered.

\subsection{Signatures in cosmological observables}

\begin{figure}
      \subfloat{\includegraphics[width=0.48\linewidth]{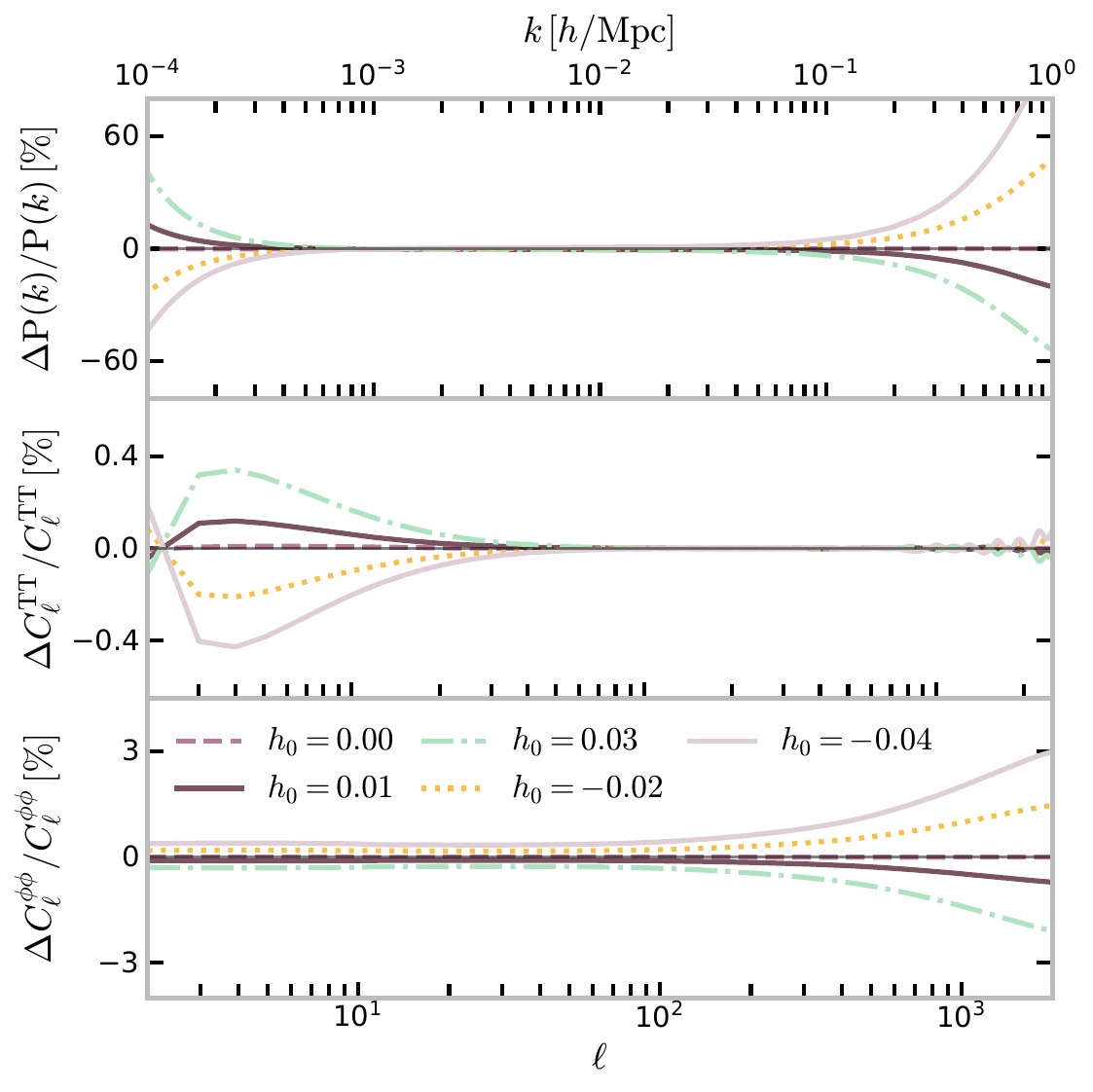}}
      \hfill
      \subfloat{\includegraphics[width=0.48\linewidth]{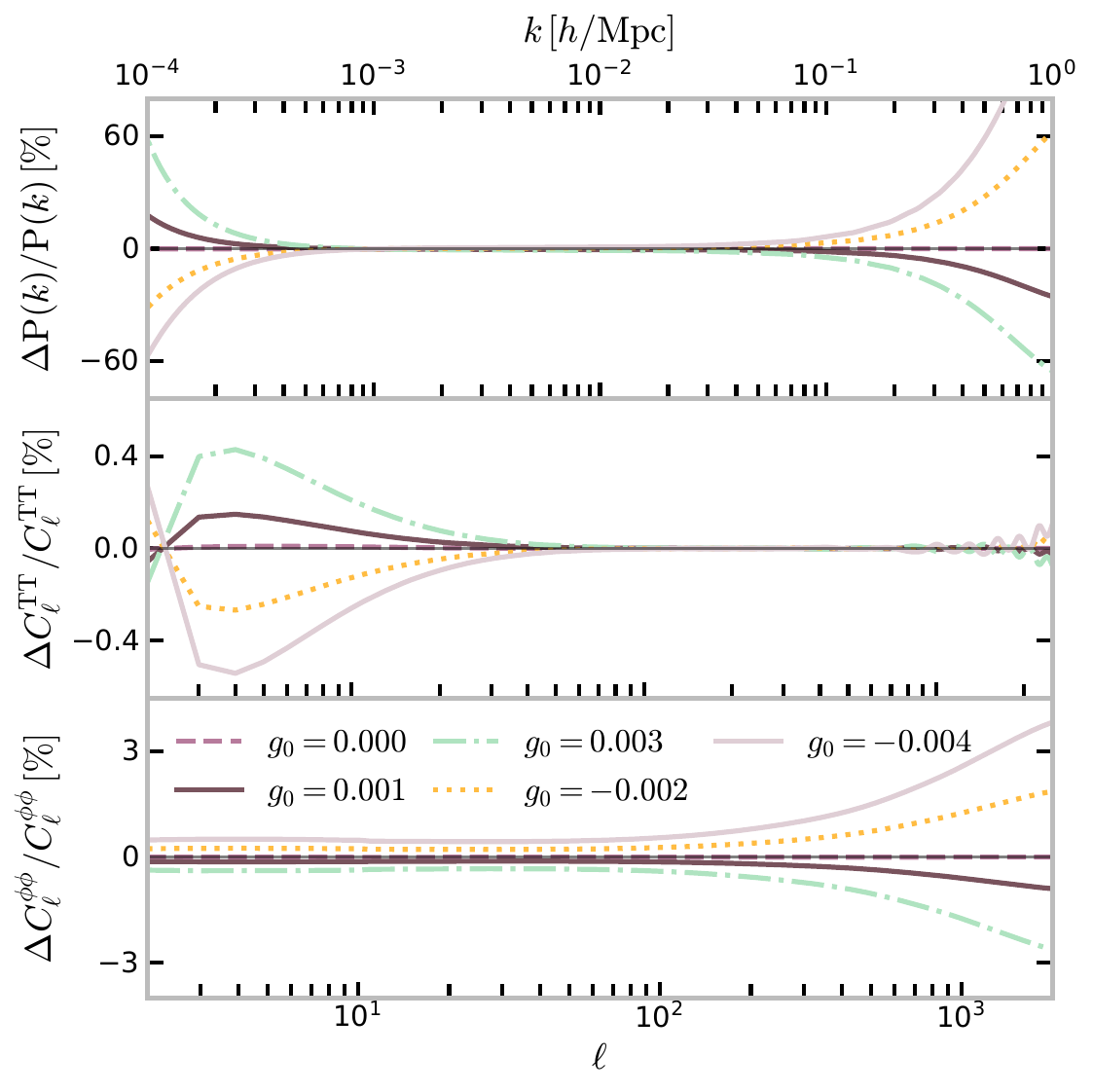}}
  \caption{Relative deviations in the matter power spectrum $P(k)$ (top panels), the CMB spectrum of temperature anisotropies $C_\ell^{TT}$ (middle panels), and the CMB lensing potential spectrum $C_\ell^{\phi\phi}$ (bottom panels), with respect to $\Lambda$CDM. The left panels show the derivative case and the right panels the algebraic case, as specified in \cref{eq:models}. The coupling parameters are varied as indicated in each panel, while the entropy perturbation $\delta s$ is defined according to \cref{eq:deltas_def}. The matter power spectrum exhibits the largest deviations, with the strongest departures on small scales driven by the entropy-induced source term in the dark matter Euler equation, while on large scales the deviations are associated with the modified evolution of the metric potentials. In contrast, the primary CMB temperature spectrum is only very weakly affected, with visible changes mainly confined to the low-multipole late integrated Sachs-Wolfe regime. The CMB lensing potential exhibits small departures arising from the line-of-sight integration of the late-time modifications of the metric potentials, effectively probing the effects of the entropy coupling.}
  \label{fig:cls}
\end{figure}

In \cref{fig:cls} we show the resulting deviations in the matter power spectrum $P(k)$ (top panels), the CMB spectrum of temperature anisotropies $C_\ell^{TT}$ (middle panels), and the CMB lensing potential $C_\ell^{\phi\phi}$ (lower panels) for the derivative (left panels) and algebraic (right panels) cases. Among these, the matter power spectrum exhibits the clearest signature of the entropy couplings. As previously discussed, the departures at small scales were expected due to the dominant effect of the entropy-induced term in the dark matter Euler equation, $k^2 Q_s \delta s$, which acts as an effective fifth force and modifies the growth of dark matter perturbations. 
The sign of the coupling determines whether the power is enhanced or suppressed relative to $\Lambda$CDM and is consistent with the trend identified for $\sigma_8$ in \cref{fig:delta_theta_DM}. In addition, as discussed in the previous sections, there can also be non-negligible effects on very large scales, associated with the modified evolution of the metric potentials through the momentum constraint. We observe that a suppression at small scales is accompanied by an enhancement at large scales, or vice versa.
Therefore, the resulting signatures in the matter power spectrum are not a simple rescaling of the amplitude, but a genuine scale-dependent effect.

In contrast, the CMB spectrum of temperature anisotropies (middle panels of \cref{fig:cls}) is only weakly affected. This is expected, since the entropy coupling leaves the background evolution unchanged and only becomes dynamically relevant at late times. As a result, the structure of the acoustic peaks of $C_\ell^{TT}$ remains essentially identical to that of $\Lambda$CDM, and no considerable deviations appear at high multipoles apart from small secondary effects induced by the modified lensing potential. The main visible changes are confined to the low-$\ell$ region since, according to the discussion in \cref{sec:large}, the entropy couplings can induce changes to the evolution of the metric potentials on large scales and at late times, leading to a subdominant integrated Sachs-Wolfe contribution on large angular scales. However, the signal remains below $1\%$, highly subdominant compared with the effects in the matter power spectrum that can go up to $\gtrsim60\%$ for the scales and coupling strengths considered. This is consistent with the fact that the entropy coupling acts as a late-time perturbative effect.

Finally, the CMB lensing potential provides a complementary probe of the same late-time physics. Since $C_\ell^{\phi\phi}$ depends on the line-of-sight integral of the Weyl potential, it is sensitive to cumulative changes in the growth of structure and in the metric perturbations over a broad range of redshifts (peaking around $z\sim2)$ and scales ($k\sim 0.01-0.1$ for current experiments).
Therefore, as illustrated in the lower panels of \cref{fig:cls}, the deviations in $C_\ell^{\phi\phi}$ trace the same late-time momentum-exchange mechanism responsible for the changes in $P(k)$. However, for the regularised entropy spectrum adopted here, the lensing signal remains moderate over the multipole range shown, consistent with the small impact on the $k$-modes that contribute most strongly to those multipoles.

\section{Discussion} \label{sec:disc} 

In this work we have introduced a consistent Lagrangian formalism for coupling the intrinsic entropy of perfect fluids to external scalar fields. Taking the former to describe dark matter and the latter to model dark energy allows for new dark sector interactions and a new realisation of pure-momentum coupling up to linear order. Notably, we find that the background evolution is unchanged from the uncoupled quintessence scenario, providing a novel way to introduce interactions that directly impact the momentum sector and matter clustering, while preserving a standard expansion history. 

To model dark matter, we introduced a barotropic and entropy-dependent equation of state (entropic-CDM), which introduces no additional non-adiabatic pressure perturbations, thus isolating the effects of the pure-entropy couplings. The new degrees of freedom associated with intrinsic entropy perturbations were found to be time-independent, consistent with past studies  \cite{Celoria:2017xos,Celoria:2017bbh}. To implement the statistical properties of these new degrees of freedom, we parametrised the spatial dependence of the entropy perturbations with a power-law spectrum, supplemented by a UV cut-off to regulate small-scale power. Adiabatic initial conditions, with vanishing initial relative isocurvature, were formally derived for the entropic-coupled system, allowing for a consistent implementation into modified Boltzmann solvers. 

Our cosmological results for the derivative and algebraic models show a significant late-time suppression or enhancement of the matter power, depending on the sign of the coupling parameters. These effects are visible directly in the dark matter perturbations and in $\sigma_8$, becoming pronounced at low redshift. The same behaviour is observed in the matter power spectrum and CMB lensing spectrum, with a suppression/enhancement on small scales. On the other hand, very large scales (close to the horizon today) are simultaneously modified by the entropy couplings, leading to a mild ISW contribution in the primary CMB spectrum. We noted that a suppression at small scales is accompanied by an enhancement at large scales, or vice versa. 
Taken together, these results indicate that the observable imprints of entropy couplings are characterised by three main features: an unchanged background evolution, late-time modifications of the dark matter velocity perturbations and structure growth, and specific scale-dependent departures from $\Lambda$CDM. Overall, the matter power spectrum (and consequently $\sigma_8$) provide the most direct signatures of the coupling. These patterns highlight the distinctive role of intrinsic entropy perturbations as external, non-dynamical sources of momentum exchange in the dark sector. In particular, the resulting deviations from $\Lambda$CDM can appear across a broad range of scales, from very large modes close to the present horizon to the small scales probed by matter clustering.

Future work will focus on confronting these models with observational data and exploring different intrinsic entropy power spectra and background potentials. For instance, studying early dark energy potentials would significantly change the time dependence of the couplings, giving rise to different observational effects. Moreover, early dark energy models are well studied in relation to the $H_0$ tension, but generically lead to unfavourable changes to matter clustering probes such as $S_8$ \cite{Poulin:2023lkg,Simon:2024jmu}. It is therefore worthwhile to investigate whether such tensions can be consistently resolved within our non-minimally coupled framework. Additionally, a thorough analysis of the stability will be conducted to ensure that viable couplings are also free from ghosts and gradient instabilities.

Alternatively, the models could be extended to include additional couplings to the particle number density, which would generically lead to energy exchange and modify the background expansion, allowing for more significant departures from $\Lambda$CDM.
Another interesting prospect is to consider coupling the intrinsic entropy of other particle species, such as neutrinos; the formulation provided in this work is completely general and could therefore equally be applied to other matter components. In that case, however, a more nuanced treatment of the relation between the particle phase space distributions and effective fluid descriptions would be required. Lastly, it would be interesting to reformulate these coupled models within an effective field theory framework, connecting them to recent developments in the literature \cite{Aoki:2025bmj}. We leave these observational and theoretical directions for future work. 

\begingroup
\makeatletter
\def\addcontentsline#1#2#3{}
\acknowledgments
\makeatother
The authors would like to thank Christian Boehmer, Lisa Mickel, Thomas Montandon, Patrick Peter and Mar\'{i}a P\'{e}rez Garrote for useful discussions and comments on the draft. We are particularly thankful to Tristan L. Smith for several discussions and contributions to an earlier draft of this work. EMT and VP are supported by funding from the European Research Council (ERC) under the European Union's HORIZON-ERC-2022 (grant agreement no. 101076865). EJ is supported by the Engineering and Physical Sciences Research Council (EPSRC) [EP/W524335/1, UKRI3030].
This article is based upon work from COST Action CA21136 Addressing observational tensions in cosmology with systematics and fundamental physics (Cosmoverse) supported by COST (European Cooperation in Science and Technology).
\endgroup

\appendix

\section{Intrinsic entropy}  \label{sec:entropy}

The discussion of entropy in cosmological scenarios is subtle, owing to the multiple, often inequivalent definitions of entropy and entropy perturbations found in the literature; see, for example, \cite{Cicoli:2021yhb,Malik:2004tf}.
To avoid confusion, we begin by considering a single perfect fluid governed by the standard thermodynamic relations \cite{Misner:1973prb,MUKHANOV1992203}. This is a discussion of the \textit{intrinsic} entropy for a cosmological fluid. In  \cref{sec:iso}, we supplement this section with a discussion of isocurvature perturbations, associated with the relative density fluctuations between distinct particle species.
For thorough reviews that cover the concepts introduced here, we refer to \cite{Kodama:1984ziu,MUKHANOV1992203}. For a more modern and introductory treatment see \cite{Malik:2008im}, while we refer to \cite{Ballesteros:2016kdx,Celoria:2017xos} for an effective field theory approach.

The entropy per particle $s$ is an intrinsic degree of freedom in the fluid description, encoding information about the internal microscopic state of the fluid — such as internal energy levels or interactions — which are often overlooked in standard cosmological analyses. Perfect fluids are by definition in a state of adiabatic flow (no shocks or heat conduction) with uniform entropy per particle $s$ along the fluid flow $u^{\mu}$, which can be expressed as
\begin{equation} \label{eq_s}
    u^{\mu} \partial_{\mu} s =0 \, .
\end{equation}
This equation follows straightforwardly from the conservation of particle number and total entropy along the fluid flow \cref{cons}. A stricter condition is that the entropy per particle is time independent and spatially uniform,
\begin{equation}
    \partial_{\mu} s =0\, ,
\end{equation}
a condition known as \textit{isentropic} flow\footnote{The study of non-isentropic fluids has attracted some interest 
in cosmological contexts \cite{Celoria:2017bbh} and can also be connected with generalised dark matter models \cite{Hu:1998kj,Kopp:2016mhm}.} \cite{Ma:1995ey,Misner:1973prb}. This implies that the energy density becomes a function of a single thermodynamic variable, typically the number density $n$.
To extend the usual cosmological discussion of matter fluids, which often neglects the effects of intrinsic entropy, we briefly outline the thermodynamic formalism for general perfect fluids that depend on both the number density and entropy per particle $\rho(n,s)$. 

We define the background equation of state $w$ and the effective sound speed squared $c_s^2$ as
\begin{align} \label{w_general}
    w &:= \frac{\bar{p}}{\bar{\rho}} = \frac{\bar{n} \bar{\rho}_{,\bar{n}}-\bar{\rho}}{
\bar{\rho}   } \, , \\ 
    c_s^2 &:= \frac{\delta p}{\delta \rho} \, , \label{c_s}
\end{align}
where an over-bar refers to the background value and a $\delta$ denotes perturbed quantities\footnote{In the main text we have expanded around a cosmological background which assumes background quantities are functions of time $\bar{q}=q(\tau)$, hence the different notation. Here we do not assume any specific background or coordinate dependence, and therefore use an over-bar notation.}.
The effective sound speed $c_s^2$ is gauge-dependent, and rest-frame corrections (denoted $\hat{c}_s^2$) can be computed in a cosmological setting, see \cref{rest_trans} \cite{Bean:2003fb,Ballesteros:2010ks}. Moreover, $c_s^2$ differs in general from the adiabatic sound speed $c_a^2$ (sometimes also called the isentropic sound speed \cite{Ellis:1998ct}), which we show shortly.

Let us now work in the thermodynamic representation, where $\rho$ and $s$ are taken as the independent variables, and other quantities such as energy density $n$ and pressure $p$ are expressed as functions of $(\rho,s)$ by assuming $\rho(n,s)$ is locally invertible \cite{Ballesteros:2016kdx}.
In the thermodynamic representation, the adiabatic sound speed squared is simply defined as
\begin{equation} \label{c_a}
    c_a^2 := \frac{\partial p}{\partial \rho}\Big|_{s}  \, .
\end{equation}
This quantity coincides with the effective sound speed $c_s^2$ for isentropic or barotropic fluids, which trivially follows from the fact that the pressure depends only on the energy density, not on entropy as an independent degree of freedom ($\partial p/\partial s =0$). In this case, $c_s^2$ can also be expressed in terms of the ratio of the time-derivatives of the density and pressure
\begin{equation}
    c_s^2|_{\textrm{barotropic}} = \frac{\delta n \bar{n} \rho_{,\bar{n}\bar{n}}}{\delta n \rho_{,\bar{n}}} = \frac{ \bar{n} \rho_{,\bar{n}\bar{n}}}{\rho_{,\bar{n}}}   =\frac{\dot{\bar{p}}}{\dot{\rho}}  = c_a^2 \, ,
\end{equation}
where an over-dot denotes the derivative with respect to a dependent variable such as time.
In addition to the adiabatic component, we define the intrinsic non-adiabatic pressure perturbation (sometimes called the \textit{intrinsic entropy perturbation} \cite{MUKHANOV1992203,Kodama:1984ziu,Malik:2008im}) in an analogous way
\begin{equation} \label{Gamma}
    \Gamma_{\textrm{int}} := \frac{\partial p}{\partial s}\Big|_{\rho} \delta s  \, .
\end{equation}
Here, $\delta s$ is the entropy perturbation per particle, and is gauge-invariant by construction \cite{Wands:2000dp,Cicoli:2021yhb}. 
The pressure perturbations can then be expressed as the sum of adiabatic and non-adiabatic parts \cite{MUKHANOV1992203}
\begin{equation} \label{p_pert_Gamma}
    \delta p = c_a^2 \delta \rho + \Gamma_{\textrm{int}} \, ,
\end{equation}
and a similar decomposition can be made for the effective sound speed 
\begin{equation}
    c_s^2 = c_a^2 + \frac{\Gamma_{\textrm{int}}}{\delta \rho} \, .
\end{equation}
This equation, \cref{p_pert_Gamma}, plays a central role in separating the physical sources of pressure perturbations, providing a fundamental relation between the adiabatic and non-adiabatic parts.

Although the entropy per particle perturbation $\delta s$ can often be rewritten in terms of the pressure and energy density perturbations through the thermodynamic relations, this is not always possible. In particular, if $\partial p/\partial s|_{\rho}=0$, then the intrinsic non-adiabatic pressure vanishes $\Gamma_{\textrm{int}}=0$ and the relation \cref{Gamma} is not invertible for $\delta s$; in this case, it follows that one cannot use \cref{p_pert_Gamma} to write $\delta s$ in terms of $\delta \rho$ and $\delta p$.
An important consequence is that vanishing non-adiabatic perturbations, $\Gamma_{\textrm{int}}=0$, does not necessarily imply $\delta s =0$. 
In fact, in barotropic fluids where $p=p(\rho)$, non-adiabatic pressure perturbations vanish identically, but the energy density may still depend non-trivially on the intrinsic entropy, $\rho=\rho(n,s)$. This means that the intrinsic entropy perturbations can be non-zero without sourcing non-adiabatic pressure, making $\delta s \neq 0$ compatible with $\Gamma = 0$. This fact appears to not be widely known and has rarely appeared in the literature, but see \cite{Ballesteros:2016kdx,Celoria:2017xos} for analogous statements in the EFT language.

To illustrate this explicitly, we consider the barotropic equation of state for an entropy-dependent fluid
\begin{equation} \label{Specific_EoS}
     \rho = m n^{1+w} \gamma(s) \implies p = w \rho \, ,
\end{equation}
where $m$ is the mass and $\gamma(s)$ is an arbitrary function of intrinsic entropy. When the background equation of state is set to zero $w=0$, we arrive at \textit{entropic-CDM}, given in \cref{CDM_EoS}. From \cref{p_pert_Gamma} we see that despite the entropy dependence in $\rho$, the sound speed is purely adiabatic and equal to the background equation of state $c_s^2 = c_a^2 = w$, and therefore the non-adiabatic pressure vanishes $\Gamma_{\textrm{int}}=0$. Similarly, the GDM parameters also all vanish for this type of fluid \cite{Hu:1998kj}.
This matches the behaviour of standard (barotropic) cosmological fluids, where the entropy dependence drops out of the equations of motion due to the absence of a non-adiabatic pressure contribution. However, in the presence of non-minimal couplings, the entropy perturbations can re-enter the cosmological dynamics, as studied in \cref{sec:coup}.

In \cref{table} we illustrate the key differences between the dark matter fluids appearing in three distinct cases: the standard cosmological scenario with a barotropic and isentropic equation of state $\rho(n)$; a fully general perfect fluid $\rho(n,s)$; and our entropic-CDM fluid, \cref{CDM_EoS}. The bottom two rows of \cref{table} derive from the cosmological entropy conservation equation, \cref{scons}.

\begin{table}[h] 
\centering
\begin{tabular}{|l|c|c|c|}
\hline
\textbf{Properties} & \ \textbf{Traditional CDM} \  & \textbf{General perfect fluid DM} & \  \textbf{``Entropic'' CDM} \ \\ 
\hline
Fluid type & Barotropic \& isentropic & Non-barotropic \& adiabatic & Barotropic \& adiabatic \\
\hline
Energy density & $\rho(n) = n m$ & $\rho(n,s)$ arbitrary & $\rho(n,s) = n m  \gamma(s)$ \\
\hline
Pressure & $p = 0$ & $p\neq 0$  & $p = 0$ \\
\hline
Effective sound speed & $c_s^2 = 0$ & $c_s^2 \neq 0$ & $c_s^2 = 0$ \\
\hline
Non-adiabatic pressure & $\Gamma_{\textrm{int}}= 0$ & $\Gamma_{\textrm{int}} \neq 0$ &  $\Gamma_{\textrm{int}} = 0$ \\
\hline
    Background entropy & $s = \textrm{const.}$ & $s = \textrm{const.}$ & $s = \textrm{const.}$ \\
\hline
Entropy perturbations &  $\delta s=0$ & $\delta s = \delta s(\vec{x})$  & $\delta s = \delta s(\vec{x})$ \\
\hline
\end{tabular}
\caption{A comparison between traditional CDM (e.g., $\Lambda$CDM, $\phi$CDM, etc.), general perfect fluid dark matter, and the entropic-CDM defined by the barotropic equation of state in \cref{CDM_EoS}. The key physical properties (pressure, effective sound speed and non-adiabatic pressure) can be uniquely derived from the energy density of each fluid model.
The final two rows (background entropy and entropy perturbations) are evaluated on cosmological backgrounds from \cref{scons}.
} \label{table}
\end{table}

\section{Isocurvature} \label{sec:iso}
In \cref{sec:entropy} we have considered only a single fluid species. In the case of multiple fluids, the relative effects between different components plays a role, which connects directly with isocurvature. Let us first define the total energy density, pressure and enthalpy of our fluids as the sum of each component \cite{Kodama:1984ziu}
\begin{equation}
    \rho = \sum_{A} \rho_{A} \, , \quad p = \sum_A p_{A} \, ,  \quad  h = \sum_{A} h_A \, , \quad h_A := \rho_A + p_A   \, . 
\end{equation}
Each component has its corresponding adiabatic sound speed $c_{a,A}^2$ given by \cref{c_a}.
The total pressure perturbation is then given by
\begin{equation}
    \delta p = c_a^2 \delta \rho + \Gamma_{\textrm{tot}} \, ,
\end{equation}
where the total (weighted) adiabatic sound speed includes contributions from each component \cite{Kodama:1984ziu}
\begin{equation}
    c_a^2 = \sum_A \frac{h_{A}}{h} c_{a,A}^2 \, .
\end{equation}
The total non-adiabatic pressure perturbation $\Gamma_{\textrm{tot}}$ can be then be decomposed into \textit{intrinsic} and \textit{relative} contributions \cite{Kodama:1984ziu}
\begin{equation} \label{Gamma_expand}
    \Gamma_{\textrm{tot}} = \sum_{A} \Gamma_{\textrm{int},A} + \Gamma_{\textrm{rel}} \, .
\end{equation}
The intrinsic part is defined for each fluid in the same way as \cref{Gamma}, while the relative non-adiabatic pressure can be written as  \cite{Kodama:1984ziu}
\begin{equation}
    \Gamma_{\textrm{rel}} = \sum_{A} \left( c_{a,A}^2 - c_a^2 \right) \delta \rho_{A}\, .
\end{equation}
Again, note that all of the different non-adiabatic pressure perturbations $\Gamma_{\textrm{tot}}$, $\Gamma_{\textrm{rel}}$ and $\Gamma_{\textrm{int},A}$ are gauge-invariant \cite{Kodama:1984ziu}. Moreover, these equations hold on arbitrary backgrounds.

Let us now define the \textit{relative entropy perturbation} (or \textit{isocurvature}\footnote{Typical examples of isocurvature include CDM–photon or neutrino–photon isocurvature modes, which can become important during recombination \cite{Malik:2002jb}. Similarly, they can be relevant in interacting models \cite{Majerotto:2009np}.}) between two species $A$ and $B$ as
\begin{equation}
S_{AB} = \frac{\delta \rho_A}{h_A} - \frac{\delta \rho_B}{h_B} \, .
\end{equation}
It should be clear that adiabatic-like initial conditions with $S_{AB}=0$ can always be chosen between any two components, and this is also true for the entropic-CDM-photon isocurvature (see \cref{sec:initial}). This can be thought of as implying a relation between the photon number perturbation $\delta n_{r}$ and the entropic-CDM number and entropy per particle perturbations $\delta n_c$ and $\delta s_c$. Indeed, for adiabatic initial conditions, we find that the intrinsic entropy perturbation enters into the initial conditions for our non-minimally coupled entropic-CDM fluid (as seen in \cref{eq:initial_cdm}).

The isocurvature relates to the relative non-adiabatic pressure in \cref{Gamma_expand} via \cite{Malik:2002jb,Malik:2004tf,Kodama:1984ziu}
\begin{equation} \label{GammaS}
    \Gamma_{\textrm{rel}} = \frac{1}{2h} \sum_{A,B} h_A h_B \left(c_{a,A}^2 - c_{a,B}^2 \right) S_{AB} \, .
\end{equation}
Note that if two fluids have the same sound speed then their isocurvature perturbations do not contribute to $\Gamma_{\textrm{rel}}$. 
The relation \cref{GammaS} holds without assuming any background field equations and is therefore valid in complete generality. One can nevertheless verify that these definitions reduce to the standard ones used in cosmology once a background FLRW metric is specified and the continuity equation is used to rewrite the terms involving $p$ in terms of $\dot{\rho}$ and $H$ \cite{Malik:2002jb,Malik:2004tf}.
What is especially important to note from \cite{Malik:2004tf} is that, in the absence of modifications to the continuity equation, vanishing initial $S_{AB}=0$ will be conserved on super-horizon scales provided the intrinsic non-adiabatic pressure \cref{Gamma} also vanishes. This is because the evolution of the isocurvature $S'_{AB}$ on large scales is sourced by intrinsic non-adiabatic pressure perturbations and the energy-exchange currents from $u^{\mu}\nabla^{\nu} T_{\mu \nu}$ (cf. Eq.~(39) of \cite{Malik:2004tf}); both of these vanish for the entropic-CDM model with pure-entropy couplings studied in this paper. This retrieves the famous result of Weinberg that adiabaticity is conserved on super-horizon scales \cite{Weinberg:2003sw}.

\section{Conservation equation}  
\label{append_cons}
Here we show that the variations of the total action in \cref{deltag,deltaJ,deltas,deltaphi,deltatheta,deltaalpha,deltaBeta,phiVar} are equivalent to the covariant conservation of the total energy-momentum tensor $\nabla^{\mu}(T_{\mu \nu} + T_{\mu \nu}^{(\phi)}+T_{\mu \nu}^{(\textrm{int})})=0$. This relies on using the fluid variable variations, \cref{deltaJ,deltas}, and the
Brown Lagrange multipliers variations, \cref{deltaphi,deltatheta}. We first use the general decomposition of a tensor into parallel and perpendicular components \cite{Ellis:1998ct}
\begin{equation}
    \nabla^{\mu} X_{\mu \nu} = h_{\nu}{}^{\lambda} \nabla^{\mu}X_{\mu \lambda} - u_{\nu} u^{\lambda} \nabla^{\mu}X_{\mu \lambda}  \, .
\end{equation}
For the (uncoupled) matter energy-momentum tensor, it is a well-known result that its projection along the fluid velocity is conserved \cite{Brown:1992kc,Boehmer:2015kta},
\begin{equation}
 u^{\lambda} \nabla^{\mu} T_{\mu \lambda} = u^{\lambda} \nabla^{\mu} \Big( (\rho+p) u_{\mu} u_{\lambda} + p g_{\mu \lambda}\Big) = 0 \, ,
\end{equation}
which follows precisely from the conservation of particle number and entropy, \cref{cons}.
Next, for the (uncoupled) scalar field energy-momentum tensor, \cref{Tscalar}, we have
\begin{equation} \label{Tphiu}
     u^{\lambda} \nabla^{\mu} T_{\mu \lambda}^{(\phi)} = u^{\lambda} \left( f_{,\phi} -\nabla_{\mu}( f_{,\mathcal{S} }  \nabla^{\mu}s) \right)  \nabla_{\lambda} \phi  \, ,
\end{equation} 
which follows from the modified Klein–Gordon equation, \cref{scalarEoM}.
Finally, for the interaction energy-momentum tensor, \cref{Tint}, we calculate the projection along the fluid flow to be
\begin{align} 
    u^{\lambda} \nabla^{\mu}T^{\textrm{int}}_{\mu \lambda} &=  u^{\lambda} \nabla^{\mu} \Big(-g_{\mu \lambda} f + 2 f_{,\mathcal{S}} \nabla_{(\mu} \phi \nabla_{\lambda)} s \Big) \nonumber \\
    &=- u^{\lambda} f_{,\mathcal{S}} \nabla_{\lambda} (\nabla^{\mu} \phi \nabla_{\mu} s) - u^{\lambda} f_{,\phi} \nabla_{\lambda} \phi +  u^{\lambda} f_{,s} \nabla_{\lambda} s +
    \nabla^{\mu}(f_{,\mathcal{S}} \nabla_{\mu} \phi) u^{\lambda} \nabla_{\lambda} s + u^{\lambda} f_{,\mathcal{S}} \nabla_{\mu} \phi \nabla^{\mu} \nabla_{\lambda}s \nonumber \\ 
    & \quad \quad \qquad \qquad + u_{\lambda} \nabla_{\mu} (f_{,\mathcal{S}} \nabla^{\mu} s ) \nabla^{\lambda}\phi + u^{\lambda} f_{,\mathcal{S}} \nabla_{\mu}s \nabla^{\mu}  \nabla_{\lambda} \phi \nonumber \\
    &= u_{\lambda} \nabla_{\mu} (f_{,\mathcal{S}} \nabla^{\mu} s ) \nabla^{\lambda}\phi -  u_{\lambda} f_{,\phi} \nabla^{\lambda} \phi +  u^{\lambda} f_{,\mathcal{S}} \Big(-\nabla_{\lambda}\nabla^{\mu}\phi \nabla_{\mu}s - \nabla^{\mu} \phi \nabla_{\lambda} \nabla_{\mu} s + \nabla_{\mu} s \nabla^{\mu} \nabla_{\lambda}\phi + \nabla_{\mu}\phi \nabla^{\mu} \nabla_{\lambda} s \Big) \nonumber \\
    &= u^{\lambda}\big( \nabla_{\mu} (f_{,\mathcal{S}} \nabla^{\mu} s ) - f_{,\phi} \big) \nabla_{\lambda}\phi \, , \label{Tintu}
\end{align}
where we have made use of $u^{\mu} \nabla_{\mu}s=0$ from \cref{entropy} and cancelled terms accordingly. Note that, up to this point, we have not needed to use all of the variational equations of motion, apart from the equations obtained from $\delta \varphi$ and $\delta \theta$ which enforce the conservation of particle number flux and of entropy along the fluid flow lines; see \cref{cons}. The remaining term in \cref{Tintu} exactly cancels \cref{Tphiu}, giving  $u^{\lambda} \nabla^{\mu} T_{\mu \lambda}^{(\textrm{total})}=0$ as expected.

Moving on to the perpendicular projection, it is now crucial to include the entropy variations, \cref{deltas}, and not to assume the isentropic condition. Following the calculations laid out by Brown \cite{Brown:1992kc}, for the perfect fluid energy-momentum tensor we straightforwardly arrive at
\begin{align}
    h_{\mu \nu} \nabla_{\lambda}T^{\lambda \nu} &= (g_{\mu \nu} + u_{\mu} u_{\nu}) \nabla_{\lambda} \big( (\rho +p) u^{\lambda} u^{\nu} + p g^{\lambda \nu} \big) \nonumber \, \\
    &=\nabla_{\lambda}\big((\rho+p) u^{\lambda} u_{\mu} + p \delta^{\lambda}_{\mu}\big) + u_{\mu} u_{\nu} \nabla_{\lambda}\big((\rho + p) u^{\lambda} u^{\nu}\big) + u_{\mu} u^{\lambda} \nabla_{\lambda} p \, , \\
    &= (\rho+p) u^{\lambda}\nabla_{\lambda}u_{\mu} + \nabla_{\mu} p + u_{\mu} u^{\lambda} \nabla_{\lambda}p \, ,
\end{align}
where we have used $u^{\lambda} \nabla_{\nu} u_{\lambda}=0$ and cancelled terms. Simplifying this term requires a bit more work, but can again be found in Ref.~\cite{Brown:1992kc}, which gives
\begin{align} \label{append_rhos}
     h_{\mu \nu} \nabla_{\lambda}T^{\lambda \nu} = n \Big[ 2 \nabla_{[\lambda}(\mu u_{\mu]}) u^{\lambda} - (\delta^{\lambda}_{\mu} + u_{\mu} u^{\lambda}) \frac{1}{n}\frac{\partial \rho}{\partial s} \nabla_{\lambda} s \Big] \,  ,
\end{align}
where $2X_{[\mu \nu]} = X_{\mu \nu} - X_{\nu \mu}$ are antisymmetry brackets and the chemical potential is $\mu = (\rho+p)/n$. Finally, let us substitute the $\mathcal{N}^{\mu}$ variation in \cref{deltaJ} into the first term, and the entropy variation in \cref{deltas} into the final term to obtain
\begin{align}
     h_{\mu \nu} \nabla_{\lambda}T^{\lambda \nu} &= -n\Big[  2 u^{\lambda}  \nabla_{[\lambda}\Big( \varphi_{,\mu]} + s \theta_{,\mu]} + \beta_{A} \alpha^{A}_{, \mu]}\Big) + \Big( u^{\lambda} \theta_{,\lambda} + \frac{1}{n} \big( \nabla_{\lambda}(f_{,\mathcal{S}} \nabla^{\lambda} \phi) - f_{,s} 
     \big) \Big) s_{,\mu}   
     \Big] \, , \\
     &= -n\Big[ u^{\lambda} \big( \nabla_{\lambda} s \nabla_{\mu} \theta -\nabla_{\mu} s \nabla_{\lambda} \theta) + u^{\lambda} \nabla_{\lambda} \theta \nabla_{\mu} s + \frac{1}{n} \big( \nabla_{\lambda}(f_{,\mathcal{S}} \nabla^{\lambda} \phi) - f_{,s} 
     \big)  \nabla_{\mu} s  \Big] \, \\
     &= \Big(f_{,s} -\nabla_{\lambda}(f_{,\mathcal{S}} \nabla^{\lambda} \phi) \Big) \nabla_{\mu}s \, , 
\end{align}
where we have used that covariant derivatives of scalars commute, and that the Lagrange multipliers $\beta_A$ and $\alpha^A$ are conserved along the fluid flow, \cref{deltaBeta,deltaalpha}. The remaining term is exactly what appears in the coupling current $J_{\mu}$ in \cref{JcoupPerp}, and can be seen as a direct consequence of coupling the matter entropy to an external field. It follows that $h_{\mu}{}^{\nu} \nabla^{\lambda} (T_{\lambda \nu} + T_{\lambda \nu}^{(\phi)}+T_{\lambda \nu}^{(\textrm{int})}) =0$ as required.

\section{Linear perturbations in the synchronous gauge} \label{sec:sync}

In this appendix we present the linear cosmological perturbation equations for the entropy-coupled dark sector model in the synchronous gauge. 
In synchronous gauge the line element is
\begin{equation}
\dd s^2 = a^2(\tau)\left[-\dd\tau^2 + \left(\delta_{ij}+h_{ij}\right)\dd x^i \dd x^j \right]\,,
\end{equation}
where the scalar perturbations of the spatial metric are parametrised in Fourier space as
\begin{equation}
h_{ij}(\vec{x},\tau) =\int \dd^3k\, e^{i\vec{k}\cdot\vec{x}} \left[ \hat{k}_i\hat{k}_j\, h(\vec{k},\tau) +\left(\hat{k}_i\hat{k}_j-\frac{1}{3}\delta_{ij}\right) 6\eta(\vec{k},\tau) \right],
\end{equation}
with $\vec{k}=k\hat{k}$.
The Newtonian gauge scalar potentials $\Phi$ and $\Psi$ are related to $h$ and $\eta$ as follows \cite{Ma:1995ey}:
\begin{eqnarray}
\Phi &=& \frac{1}{2k^2} \left[ h'' + 6\eta'' + \mathcal{H}\left(h'+6\eta'\right) \right], \\
\Psi &=& \eta - \frac{\mathcal{H}}{2k^2} \left(h'+6\eta'\right),
\end{eqnarray}
where primes denote derivatives with respect to conformal time $\tau$, and $\mathcal{H}\equiv a'/a$.

Using the synchronous–gauge metric variables $h$ and $\eta$, the Einstein equations corresponding to \cref{EFE1,EFE2,EFE3,EFE4} can be written as
\begin{align}
k^2 \eta - \frac{1}{2}\mathcal{H} h' &=  4 \pi G \left( a^2 \sum_{c,b,r} \delta \rho_I + a^2 \hat{V}_{,\phi}\,\delta\phi + a^2 g_{,s}\,\delta s + \phi' \delta\phi' \right) \,, \label{EFE1_sync}\\
h'' + 2\mathcal{H} h' - 2k^2 \eta &= -8 \pi G \left( a^2 \sum_{c,b,r} \delta p_I - a^2 \hat{V}_{,\phi}\,\delta\phi - a^2 g_{,s}\,\delta s + \phi' \delta\phi' \right) \,, \label{EFE2_sync}\\
k^2 \eta' &= 4 \pi G \left( a^2 \sum_{c,b,r} (\rho_I + p_I)\theta_I + k^2 \phi' \big(h_0 \delta s + \delta\phi \big) \right) \, , \label{EFE3_sync} \\
h'' + 6 \eta'' + 2 \mathcal{H}(h' + 6 \eta') - 2 k^2 \eta &= -24 \pi G a^2 \sum_{c,b,r} (\rho_I +p_I) \sigma_I \, .
\end{align}
For a generic uncoupled fluid species $I$, the continuity and Euler equations, \cref{cont1,Eul1}, in the synchronous gauge are
\begin{eqnarray}
\delta_I' + 3\mathcal{H} \left(\frac{\delta p_I}{\delta\rho_I}-w_I\right)\delta_I + (1+w_I)\left(\theta_I+\frac{h'}{2}\right) &=& 0 \, , \\
\theta_I' + \left[\mathcal{H}(1-3w_I) + \frac{w_I'}{1+w_I}\right]\theta_I - \frac{\delta p_I}{\delta\rho_I} \frac{k^2}{1+w_I}\delta_I + k^2 \sigma_I &=& 0 \, ,  \label{delete_label}
\end{eqnarray}
where $w_I = p_I/\rho_I$ is the background equation of state and we have included the anisotropic shear term $\sigma_I$.

We now specialise to our entropic-CDM fluid \cref{CDM_EoS} with vanishing equation of state $w_c = 0$, shear $\sigma_c=0$ and pressure perturbations $\delta p_c = 0$.
From the purely-perpendicular covariant coupling current, \cref{JcoupPerp}, there is no energy transfer at all orders. The continuity equation for cold dark matter, \cref{DM_delta}, therefore takes the standard form
\begin{equation}
\delta_c' + \theta_c + \frac{h'}{2} = 0 \,.
\end{equation}
The entropy dependence appears only through a pure-momentum transfer in the Euler equation, \cref{DM_theta}, which reads
\begin{equation}
\theta_c' + \mathcal{H}\theta_c = - \frac{k^2}{\rho_c} \left( g_{,s} -h_0 V_{,\phi} \right)\delta s \,,
\end{equation}
where $g(s,\phi)$ is the entropy algebraic coupling function and $h_0$ denotes the amplitude of the derivative entropy-scalar coupling $h(\mathcal{S})$ (which is unrelated to the metric perturbation $h$), according to \cref{eq:f_function}. 
Also recall that the intrinsic entropy is a gauge-independent degree of freedom and conserved along the dark matter flow, so that $\delta s$ is non-dynamical and fully specified by its primordial spectrum.

Finally, the scalar field perturbation equation, \cref{eq:pert_kg}, is
\begin{equation}
\delta\phi'' + 2\mathcal{H}\delta\phi' + \left(a^2 V_{,\phi\phi} + k^2\right)\delta\phi + \frac{h'}{2}\phi' = - \left( a^2 g_{,s\phi} + k^2 h_0 \right) \delta s \,.
\end{equation}
It is worth noting that the synchronous gauge is defined as the frame which is comoving with cold dark matter in the absence of interactions. When the entropy coupling
vanishes, $h_0=g_{,s}=0$, and for initial conditions satisfying $\theta_c(\tau_i)=0$, the velocity divergence of cold dark matter remains zero at all times. However, in the entropy-coupled model, the Euler equation has an additional scale-dependent source term proportional to $\delta s$, generating a pure-momentum exchange that leaves the background evolution and energy conservation unchanged while modifying the growth of structure. For this reason, and for numerical purposes, we defined this gauge in terms of a very subdominant uncoupled dark matter component.

\section{Entropy initial conditions} \label{sec:initial}

In this section we derive the early-time behaviour of the perturbations in the synchronous gauge for our dark matter fluid and scalar field dark energy in the presence of the non-minimal entropy coupling introduced in \cref{sec:entropic}, following the notation and derivation of Ref.~\cite{Ballesteros:2010ks}.
We work in the synchronous gauge during radiation domination,
\begin{equation}
a(\tau) = \alpha \tau, \qquad \mathcal{H} = \frac{1}{\tau} \, ,
\end{equation}
and consider super-horizon modes $k\tau \ll 1$ and $\theta_c (k\tau_i)=0$ by definition.

The adiabatic growing metric solution is
\begin{equation}
h = C (k\tau)^2 \, ,
\qquad 
h' = 2 C k^2 \tau \, .
\end{equation}
For the adiabatic growing mode, we do not excite an independent isocurvature mode and can therefore set the homogeneous scalar perturbation to zero.  Nevertheless, the entropy coupling introduces a source term in the perturbed Klein-Gordon equation proportional to $\delta s$, which generates a non-vanishing contribution. The initial conditions for $\delta \phi$ will be obtained by expanding the regular solution in powers of $(k\tau)$.

First, we look at initial conditions for the density contrast and velocity perturbation of the entropic-CDM fluid. Consistently with adiabatic growing modes, we assume the lowest-order behaviour in $k \tau$ for $\delta_c$ and $\theta_c$ and consider a possible contribution at next to leading order:
\begin{equation}
\label{eq:ansatz}
\delta_c = A (k\tau)^2 + E (k\tau)^3\, ,
\qquad
\delta_c' = 2 A k^2 \tau + 3 E k^3 \tau^2 \, ,
\qquad
\theta_c = B (k\tau)^3 + F (k \tau)^4 \, ,
\qquad
\theta_c' = 3 B k^3 \tau^2 + 4 F k^4 \tau^3\, .
\end{equation} 
The equations in the synchronous gauge are
\begin{align}
\label{eq:rd_eqs}
\delta_c' + \theta_c + \frac{h'}{2}  &= 0\, ,\\
\theta_c' + \mathcal{H}\theta_c &= - k^2 D\,\delta s\,\tau^{3} \quad \text{with}\quad  D=\frac{g_{,s} - h_0 \hat{V}_{,\phi}}{\rho_{c,0} \tau_0^{3}}\, ,
\end{align}
where we used $\rho_c \approx \rho_{c,0} \left(\tau_0/\tau\right)^3$, with $\tau_0$ being some reference time in radiation domination. We approximate $D$ to be a constant at this epoch since the field is slowly rolling in the early Universe ($g_{,s}\approx \text{const.}$ and $h_0 \hat{V}_{,\phi}\approx \text{const.}$).

Replacing the ansatz in \cref{eq:ansatz} into \cref{eq:rd_eqs} and matching the powers on each side  gives 
\begin{equation}
A = -\frac{C}{2} \, , \quad B = 0\, , \quad E =  0\, , \quad F = -\frac{D\,\delta s}{5k^2} \, . 
\end{equation}
Thus, the initial conditions for $\delta_c$ and $\theta_c$ are
\begin{equation} \label{eq:initial_cdm}
    \delta_c(\tau)=-\frac{C}{2}(k\tau)^2\, , \quad
\theta_c(\tau)=-\frac{D\,\delta s}{5k^2}\,(k\tau)^4\, .
\end{equation}
Because the entropy term scales differently in time, it generates a new independent power. 
The entropy source generates the $F(k\tau)^4$ correction in the solution of $\theta_c$, while $\delta_c$ retains its standard adiabatic growing solution at this order.

We now move on to the scalar field initial conditions.
The perturbed Klein-Gordon equation in the synchronous gauge is
\begin{equation}
\delta\phi'' + 2\mathcal{H}\delta\phi'
+ (k^2 + a^2 \hat{V}_{,\phi\phi})\delta\phi
+ \frac{1}{2}h'\phi'
= -(a^2 g_{,s\phi} + k^2 h_0)\delta s \, .
\end{equation}
Since $\phi'=\mathcal{O}(\tau^3)$, the metric-drag term is higher order and the leading source is $(a^2 g_{,s\phi} + k^2 h_0)\delta s$.
Expanding $\delta \phi$ in powers of $(k \tau)$
\begin{align}
\delta\phi &= \delta\phi_0 + \delta\phi_1(k\tau) + \delta\phi_2(k\tau)^2 + \delta\phi_3(k\tau)^3 + \delta\phi_4(k\tau)^4 \, ,\\
\delta\phi' &= \delta\phi_1 k + 2\delta\phi_2 k (k\tau) + 3\delta\phi_3 k (k\tau)^2 + 4\delta\phi_4 k (k\tau)^3 \, ,\\
\delta\phi'' &=  2\delta\phi_2 k^2 + 6 \delta\phi_3 k^2 (k\tau) + 12\delta\phi_4 k^2 (k\tau)^2 \, .
\end{align}
Replacing this expansion in the Klein-Gordon equation and matching orders gives:
\begin{equation}
\delta\phi_1=\delta\phi_3=0,
\qquad
6\delta\phi_2 = -h_0\delta s \qquad
20k^4 \delta\phi_4 = - \alpha^2 g_{,s\phi} \delta s \, .
\end{equation}
Thus, the first non-trivial contributions from the entropy source term lead to
\begin{align}
\delta\phi(\tau)
&= -\frac{h_0}{6}\delta s\,(k\tau)^2 - \frac{\alpha^2 g_{,s\phi} \delta s}{20 k^4}\,(k\tau)^4 +  \mathcal{O}\left((k\tau)^5\right)\, , \\ 
\delta\phi'(\tau) 
&= -\frac{h_0}{3}\delta s\,k\,(k\tau) - \frac{\alpha^2 g_{,s\phi} \delta s}{5 k^3}\, (k\tau)^3 + \mathcal{O}\left((k\tau)^4\right) \, .
\end{align}

\begin{figure}
      \subfloat{\includegraphics[height=0.35\linewidth]{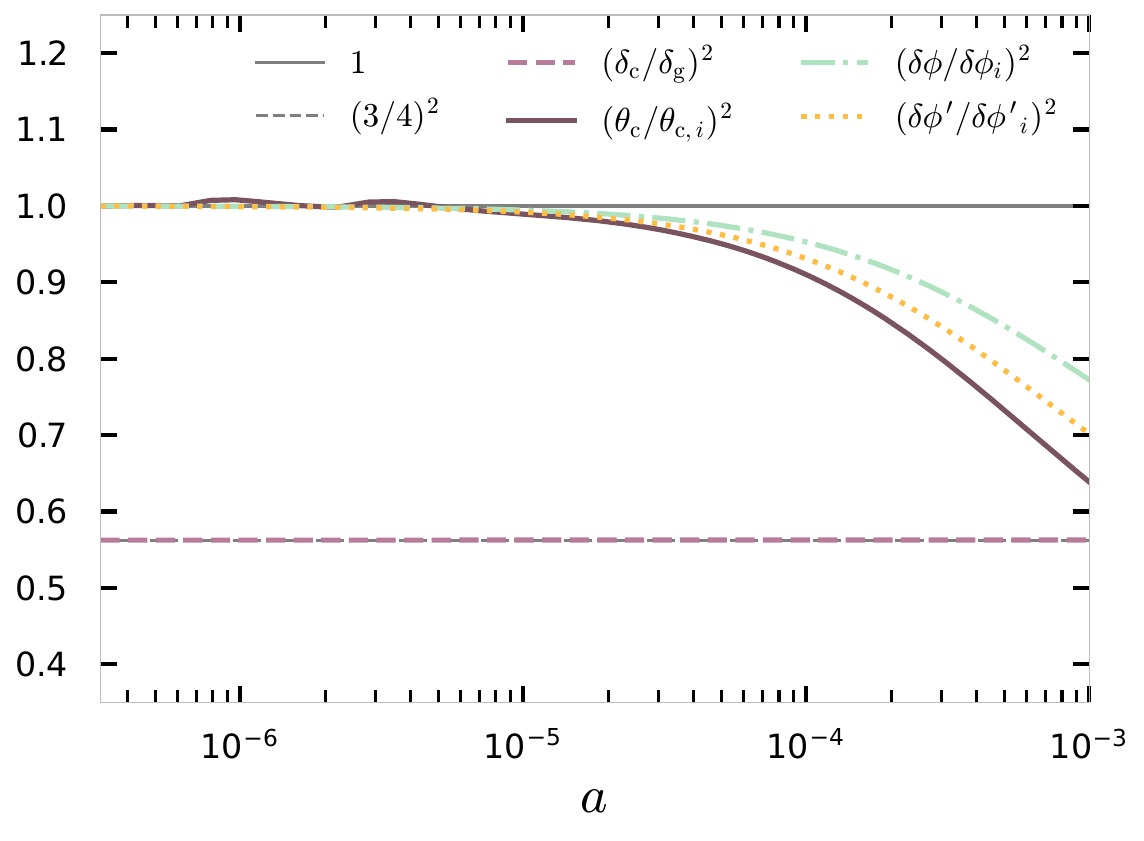}}
      \qquad
      \subfloat{\includegraphics[height=0.35\linewidth]{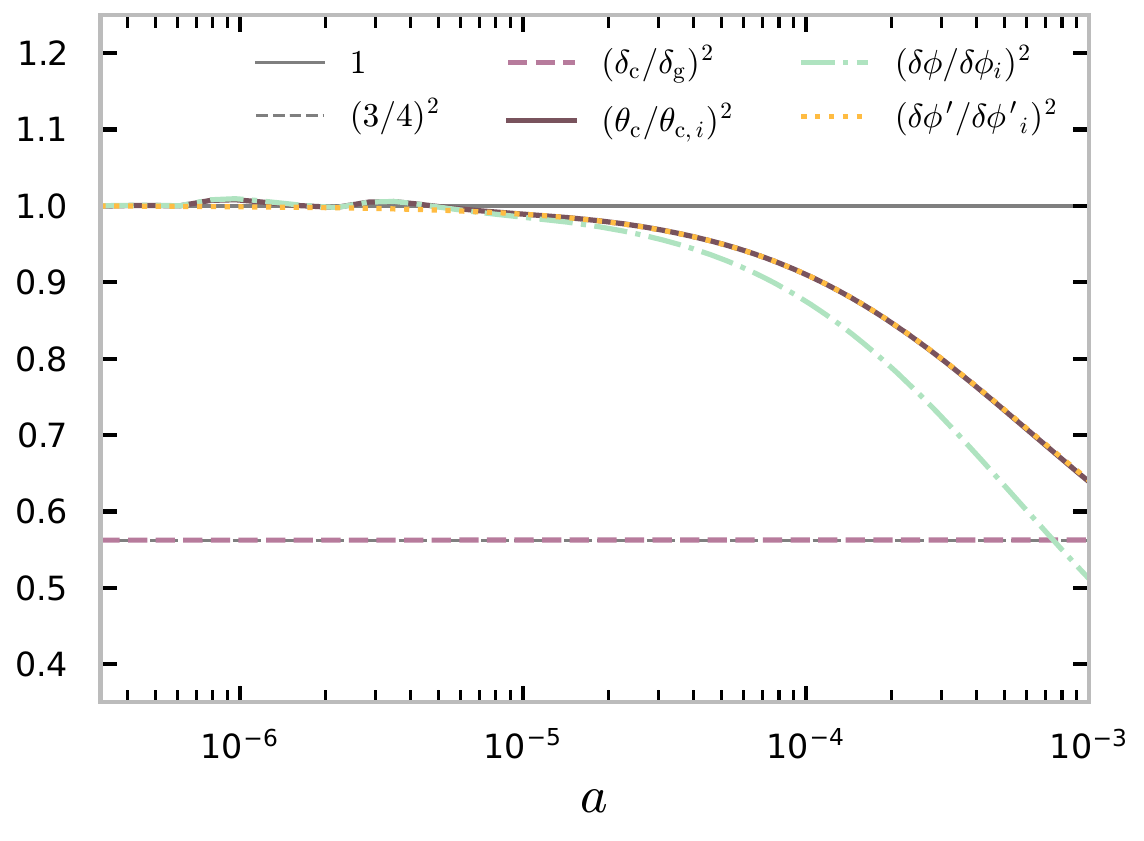}}
\caption{Early-time scaling of the numerical solutions for $\delta_c$, $\theta_c$, $\delta\phi$, and $\delta\phi'$ at $k=10^{-4}\,\mathrm{Mpc}^{-1}$. All quantities are normalised to their expected leading-order super-horizon adiabatic initial conditions, showing agreement with the analytic expansion derived in \cref{sec:initial}. The entropy perturbation $\delta s$ is parameterised as in \cref{eq:deltas_def}. The left panel corresponds to the derivative case with $h_0=0.01$, and the right panel to the algebraic case with $g_0=0.001$, as defined in \cref{eq:models}.}
  \label{fig:condinis}
\end{figure}

We implement these initial condition for the entropy coupled cold dark matter and the dark energy scalar field in the perturbation module of \texttt{CLASS}. 
In \cref{fig:condinis} we verify the validity of the analytic expansion derived in this appendix, according to the numerical setting described in \cref{sec:obs}. For both the derivative and algebraic couplings, the numerical solutions for $\delta_c$, $\theta_c$, $\delta\phi$, and $\delta\phi'$ approach the expected leading-order super-horizon scalings once each quantity is normalised to its corresponding analytic initial condition. This confirms that the modified initial conditions implemented in \texttt{CLASS} correctly reproduce the entropy-sourced contributions derived above. 

\nocite{DataAvailibility}
\bibliography{bib}

\end{document}